\PassOptionsToPackage{table}{xcolor}
\documentclass[fleqn,10pt]{wlscirep}
\usepackage[utf8]{inputenc}
\usepackage[T1]{fontenc}

\usepackage{times}  
\usepackage{helvet}  
\usepackage{courier}  
\usepackage{subcaption}
\usepackage{soul}
\usepackage{multirow}
\usepackage{ulem}
\usepackage{graphicx}

\let\oldtextsf\textsf

\renewcommand{\textsf}[1]{{\fontsize{8.5}{10}\selectfont\oldtextsf{#1}}}

\newcommand{\ita}[1]{#1}

\title{Unmasking Parkinson's Disease with Smiles: An AI-enabled Screening Framework}

\author[1,$\dagger$,*]{Tariq Adnan}
\author[1,$\dagger$]{Md Saiful Islam}
\author[1]{Sangwu Lee}
\author[1]{Wasifur Rahman}
\author[2]{Sutapa Dey Tithi}
\author[2]{Kazi Noshin}
\author[3]{Imran Sarker}
\author[2]{M Saifur Rahman}
\author[4]{Ruth B. Schneider}
\author[4]{Jamie L. Adams}
\author[4]{E Ray Dorsey}
\author[1]{Ehsan Hoque}

\affil[1]{Department of Computer Science, University of Rochester, Rochester, New York, USA}
\affil[2]{Department of Computer Science and Engineering, Bangladesh University of Engineering and Technology, Dhaka, Bangladesh}
\affil[3]{Department of Neurology, National Institute of Neurosciences \& Hospital, Dhaka, Bangladesh}
\affil[4]{Department of Neurology, University of Rochester Medical Center, Rochester, New York, USA}
\affil[*]{tadnan@ur.rochester.edu}
\affil[$\dagger$]{These authors contributed equally to this work.}

\begin{abstract}
\textbf{Background} Parkinson's disease (PD) diagnosis is challenging due to the limited availability of reliable biomarkers and insufficient access to clinical care. Here, we present an efficient and accessible PD screening method by leveraging AI-driven models enabled by the largest video dataset of facial expressions from 1,059 unique participants. 
This dataset includes 256 individuals with PD, 165 of whom were clinically diagnosed and 91 were self-reported.\\

\noindent \textbf{Methods} Participants used webcams to record themselves mimicking three facial expressions (smile, disgust, and surprise) from diverse sources encompassing their homes across multiple countries, a US clinic, and a PD wellness center in the US. Facial landmarks are automatically tracked from the recordings to extract features related to hypomimia, a prominent PD symptom characterized by reduced facial expressions. Machine learning algorithms are trained on these features to distinguish between individuals with and without PD. The model was tested for generalizability on external (unseen during training) test videos collected from a US clinic and from Bangladesh.\\

\noindent\textbf{Findings} An ensemble of machine learning models trained on smile videos achieved an accuracy of $87.9 \pm 0.1\%$ ($95\%$ Confidence Interval) with an AUROC of $89.3 \pm 0.3\%$ as evaluated on held-out data (using $k$-fold cross-validation). In external test settings, the ensemble model achieved $79.8 \pm 0.6\%$ accuracy with $81.9 \pm 0.3\%$ AUROC on the clinical test set and $84.9 \pm 0.4\%$ accuracy with $81.2 \pm 0.6\%$ AUROC on participants from Bangladesh. In every setting, the model was free from detectable bias across sex and ethnic subgroups, except on the cohorts from Bangladesh, where the model performed significantly better for female participants compared to the males.\\

\noindent\textbf{Conclusion} Smiling videos can effectively differentiate between individuals with and without PD, offering a potentially easy, accessible, and cost-efficient way to screen for PD, especially when a clinical diagnosis is difficult to access.\\

\noindent\textbf{Funding} US National Institutes of Health (NIH), Gordon and Betty Moore Foundation, and Google.
\end{abstract}
\begin{document}

\maketitle
%
%
\thispagestyle{empty}

\vspace*{-1cm}
\section{Introduction}
Parkinson's disease (PD) is a progressive neurodegenerative disease and the fastest-growing neurological disorder in the world~\cite{dorsey2018emerging}. PD is typically diagnosed based on the presence of characteristic motor symptoms and signs, such as bradykinesia, rigidity, and tremor. However, by the time these characteristic motor symptoms manifest themselves, brain pathology is already advanced with loss of approximately $60\%$ of substantia nigra dopaminergic neurons~\cite{dauer2003parkinson}. Moreover, establishing an accurate clinical diagnosis can be challenging for individuals, especially in underprivileged areas due to limited access 
to the
trained neurologists. 
In many regions of the world, including the U.S., geographic distribution of neurologists is uneven~\cite{lin2021geographic}. 
Study conducted in Mississippi~\cite{schoenberg1985prevalence} indicates that up to $40\%$ of individuals with PD remain undiagnosed of the condition.
In Bangladesh, where we conducted our study, there were only 86 neurologists for over 140 million people in 2014~\cite{chowdhury2014pattern}. A recent study highlighted the extreme scarcity of neurologists in Africa, with ten nations having no neurologists at all~\cite{kissani2022does}. This acute shortage underscores the critical need for alternative, accessible, and innovative approaches to PD screening and diagnosis that can reach broader populations. A reliable, easily accessible diagnostic marker for PD is urgently needed~\cite{yang2022artificial} to support early, accurate diagnosis and enable earlier interventions.  

A characteristic often associated with PD is hypomimia, which refers to a reduction in facial expressions. People with PD may experience hypomimia due to a decrease in dopamine synthesis critical for facial expression, caused by the loss of certain neurons~\cite{makinen2019individual}. While hypomimia is not exclusive to PD, it is often considered an early and sensitive biomarker for the disease and can be utilized for early screening~\cite{abrami2021automated, maycas2021hypomimia}. Recent advancements in artificial intelligence (AI) have made it possible to identify individuals at risk of PD by analyzing facial expressions from videos~\cite{su2021detection,
novotny2022automated}. This approach offers several advantages, including the ability to conduct screenings remotely using smartphones or computers, making it accessible to a wider population. Given that nearly half of private households worldwide had desktop or laptop computers in 2019~\cite{statista_smartphone_users}, and the current global usage of smartphones (7.2 billions~\cite{taylor2023number}), this method provides a simple, convenient, and accessible screening option for anyone, regardless of location. 

Collecting data from many representative participants is difficult, as going door to door to participants' houses across the country (and potentially the world) is infeasible. This logistical difficulty has led previous studies to rely on smaller, often homogeneous datasets collected in controlled lab settings~\cite{su2021detection, abrami2021automated, novotny2022automated, bandini2017analysis}, which limits the generalizability of their findings. To address these limitations, we utilized PARK, a web-based framework designed for global data collection~\cite{langevin2019park}. Through strategic collaborations with several institutions -- including the Udall Center at the University of Rochester Medical Center, InMotion (a PD wellness facility in Ohio), the National Institute of Neurosciences \& Hospital (NINS) in Bangladesh, and Bangladesh University of Engineering and Technology -- we have gathered a diverse dataset. Enabled by 10M USD funding from the National Institutes of Health (NIH), this dataset includes facial expression (i.e., smile, surprise, and disgust) videos recorded using various devices in the home and clinical settings. Overall, we have compiled videos from 1,059 participants, including 256 individuals with PD. This dataset represents one of the largest collections of its kind for PD screening and encompasses a broad range of ages and demographics.

\begin{figure*}[!htbp]
    \centering
    \includegraphics[width=\linewidth]{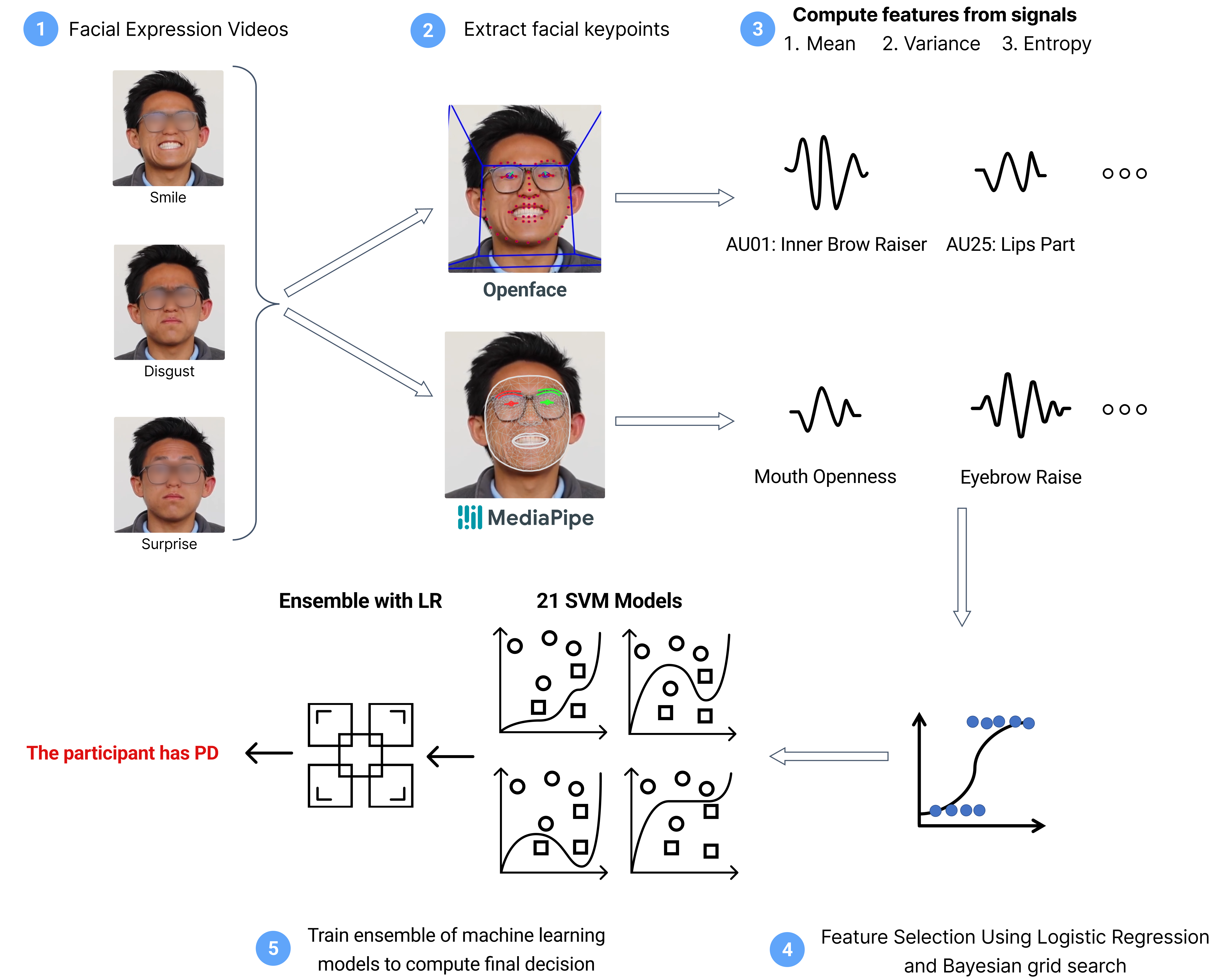}
    \caption{\textbf{A brief overview of the AI-based system for classifying individuals with and without PD.} Anyone can record their smile expressions in front of a computer webcam. The system extracts facial keypoints for each frame using Openface and Mediapipe tools, and then computes features by summarizing the temporal dimensions with statistical aggregates. 
    Top-$n$ features are then fed to an ensemble of models which finally decides whether the participant has PD or not.
    }
    \label{fig:overview}
\end{figure*}

Considering the universal nature of smile, a language-independent facial expression that can be performed by anyone with simple instructions, we explored the potential of using smiling videos to differentiate individuals with and without PD. Leveraging advanced computer vision tools, such as Mediapipe~\cite{lugaresi2019mediapipe} and Openface~\cite{openface}, we extracted objectively quantifying features of hypomimia highly aligned with the Movement Disorder Society-Sponsored Revision of the Unified Parkinson's disease Rating Scale (MDS-UPDRS)~\cite{goetz2008movement}. Our PD screening tool was tested on external cohorts in Bangladesh 
(a densely populated country from South Asia, and South Asia represents $25.2\%$ of the global population)
and the United States (representing Northern America, $4.7\%$ of the global population) to further validate its effectiveness across diverse demographics and contrasting socio-economic backgrounds. We present an overview of our proposed framework in Figure \ref{fig:overview}.

\subsection*{Implications of the study}
Smiling videos can potentially be utilized as a reliable digital biomarker for screening PD. With the prevalence of smartphones or computers, PD could be passively screened with increasingly common devices. The technology is convenient, accessible, and cost efficient -- especially suitable for geographical areas where neurological care is hard to access. Further research to assess the utility of such an approach is needed.

\section{Related Works}
The automatic detection of PD using facial expression videos has been explored in various studies. This section reviews the existing literature on this topic, highlighting the methods, datasets, and key findings, with their limitations. Su et al. conducted a study to detect hypomimia in PD patients using smile videos \cite{su2021detection}. They collected videos in a controlled environment with a fixed setup, including a solid color background and ambient lighting. The participants maintained a fixed facial position with no head movements, producing one neutral and one smile expression. The study identified $68$ facial keypoints to compute facial geometric features and used the Histogram of Oriented Gradients (HOG)~\cite{dalal2005histograms,deniz2011face} algorithm for textured features. The dataset consisted of $47$ PD patients and $39$ controls, all clinically validated. The authors reported an F1 score of $99.94\%$ and a recall of $100\%$, although concerns about overfitting due to the small dataset and lack of model development details were raised. The dataset and code were not made publicly available, limiting the replicability of the study. Bandini et al. analyzed facial expressions in PD patients using video-based automatic methods \cite{bandini2017analysis}. They captured multiple basic expressions (happiness, anger, disgust, sadness, and neutral) using a Microsoft Kinect~\cite{zhang2012microsoft} sensor in a clinical setup. The study involved $17$ PD patients and $17$ controls, all clinically validated. They detected facial expressions using a trained Support Vector Machine (SVM) model. While the model performed well on control videos, it was less accurate for PD patients, often misclassifying facial expressions as neutral. The small dataset and the focus on analyzing feature differences rather than PD classification were noted as limitations.
Abrami et al. proposed a computer vision system to assess hypomimia in PD using Convolutional Neural Networks (CNNs) trained on video frames \cite{abrami2021automated}. They collected $3425$ videos of $1595$ individuals from the YouTube Faces Database for the control group and $107$ interview videos of self-reported PD participants. The study reported a $70\%$ accuracy in detecting hypomimia and claimed $100\%$ accuracy in a subset of videos from a single actor before PD diagnosis. The main limitations were the use of unvalidated control videos and the reliance on interview videos of self-reported PD patients from YouTube, leading to potential unreliable labeling issues.
Novotny et al. conducted a study on facial bradykinesia in de-novo PD patients using a large video dataset \cite{novotny2022automated}. They recruited $91$ PD participants and $75$ controls, all of Czech nationality, and captured videos in a clinical environment. This study had the largest dataset among the reviewed works but faced limitations regarding the generalizability of the findings due to the homogeneity of the participants' demographics. 

Despite the promising results reported in these studies, several limitations are common across them. The small and homogeneous datasets, use of constrained and fixed recording environment, limited generalizability of the findings, and lack of publicly available code and datasets are significant challenges. To address these issues, we built our work on a large-scale dataset diverse in demographic properties like age, sex, and ethnicity. We also externally validated our model to ensure generalizability.
Furthermore, we will publish our code and feature set upon acceptance of our manuscript, enabling other researchers to validate and build upon our work.


\section{Method}

\subsection*{Study design} This study invited participation from four distinct settings, where a 
participant may seek care for Parkinson's disease -- (i) from participants' homes, located anywhere in the world (\texttt{Home-Global} cohort), (ii) from a US clinic (\texttt{Clinic} cohort), (iii) from a PD care facility located in the US (\texttt{PD Care Facility} cohort), and (iv) from participants' homes, located in Bangladesh, representing a low-income country in South Asia (\texttt{Home-BD} cohort). We used a web-based framework named PARK~\cite{langevin2019park} for collecting videos of three facial expressions (i.e., disgust, smile, and surprise). For each expression, participants were asked to mimic it as expressively as possible, sustain the expression for a few seconds, and then revert to a neutral expression. This was repeated three times in each recorded video, and the three expressions were recorded as three different videos. In addition to the video recordings, PARK collected demographic information about the participants -- age, sex, ethnicity, and nationality. Note that, while extracting features, we only utilized the video recordings and did not incorporate the demographic information. Instead, we reserved the demographic data for post-modeling analysis, such as bias analysis to identify underperforming demographic subgroups. Including demographic information in the modeling purpose might also restrict the global usability of the model, as the collection of such data is subject to strict regulations in many countries, including several in Europe, which often prohibit or limit the collection of participants' racial and ethnic information~\cite{simon2007ethnic,simon2015choice}.
To ensure data quality, we provided instructional videos showing how to complete the task. This was supplemented by guidelines encouraging participants to record themselves in well-lit environments, against distinguishable backgrounds, and to position their full faces within the recording frame. Details about recruitment and demographic information for each cohort are provided in the supplementary materials (see Supplementary Notes 1 and 5).

This study was approved by the Institutional Review Board (IRB) of the University of Rochester, the University of Rochester Medical Center, and Bangladesh University of Engineering \& Technology. All the experimental procedures complied with the approved study protocol. Given the predominantly remote administration of this study, written consent from participants was not obtained. However, informed consent was duly collected electronically from the participants, authorizing the use of their data for analysis and their photos in relevant publications.

\subsection*{Dataset}
Our video dataset is sourced from $1059$ participants, comprising $256$ diagnosed with PD and $803$ without the condition. The \texttt{Home-Global} cohort consists of $693$ global participants, however, dominated by US residents ($89.5\%$) and participants without PD ($88.9\%$). Note that the PD participants in this cohort are self-reported. The \texttt{Clinic} cohort consists of $75$ participants with $47$ clinically diagnosed with PD. For the \texttt{PD Care Facility} cohort, the PD samples are collected from the clients ($n = 118$), who are clinically diagnosed to have PD, while the non-PD samples are collected from their caregivers ($n = 24$). Although the \texttt{Home-BD} cohort represents a contrasting socio-economic background, they are primarily dominated by participants without PD -- only $14$ out of $149$ participants in this cohort self-reported to have PD. 
$56$ individuals received their diagnosis within five years of data collection, while $65$ were diagnosed more than five years prior. Unfortunately, we could not collect disease duration information from $135$ participants. 
A brief overview of the participant demography with the statistics of disease duration is provided in Table \ref{tab:demographics}. 

\begin{table}[!htbp]
\centering
\caption{\textbf{Dataset Demographics}. Demographic information of the (a) full participating cohorts, (b) training and test set participants.}
\subcaptionbox{\label{tab:demographics}}{
\resizebox{0.65\columnwidth}{!}{
\begin{tabular}{lllll}
\toprule
\multicolumn{2}{l}{\textbf{Characteristics}} & \textbf{With PD} & \textbf{Without PD} & \textbf{Total} \\ \hline
\multicolumn{2}{l}{Number of Participants, n (\%) } & $256\ (24.2\%)$ & $803\ (75.8\%)$ & \textbf{1059 (100\%)} \\ \hline
\multicolumn{5}{l}{Sex, n (\%)} \\
& Male & $150\ (58.6\%)$ & $361\ (44.9\%)$ & $\mathbf{511\ (48.3\%)}$ \\ 
& Female & $106\ (41.4\%)$ & $442\ (55.1\%)$ & $\mathbf{548\ (51.7\%)}$ \\ 
\midrule
\multicolumn{5}{l}{Age in years (range: $18 - 93$, mean: $58.9$), n (\%)} \\
& \textless{}$20$ & $0\ (0.0\%)$ & $43\ (5.4\%)$ & $\mathbf{43\ (4.1\%)}$ \\
& $20-39$ & $4\ (1.6\%)$ & $96\ (12.0\%)$ & $\mathbf{100\ (9.4\%)}$ \\
& $40-59$ & $39\ (15.2\%)$ & $216\ (26.8\%)$ & $\mathbf{255\ (24.1\%)}$ \\
& $60-79$ & $198\ (77.3\%)$ & $438\ (54.6\%)$ & $\mathbf{636\ (60.0\%)}$ \\
& \textgreater{}$=80$ & $15\ (5.9\%)$ & $10\ (1.2\%)$ & $\mathbf{25\ (2.4\%)}$ \\
\midrule
\multicolumn{5}{l}{Race, n (\%)} \\
 & White  & $116\ (45.3\%)$ & $531\ (66.1\%)$ & $\mathbf{647\ (61.1\%)}$ \\
 & {Asian}  & $15\ (5.8\%)$ & $187\ (23.3\%)$ & $\mathbf{202\ (19.1\%)}$ \\
 & {Black or African American} & $3\ (1.2\%)$ & $45 (5.6\%)$ & $\mathbf{48\ (4.5\%)}$ \\
 & {{American} {Indian or} {Alaska} {Native} } & $1\ (0.4\%)$ & $4\ (0.5\%)$ & $\mathbf{5\ (0.5\%)}$ \\
& Others  & $3\ (1.2\%)$ & $12\ (1.5\%)$ & $\mathbf{15\ (1.4\%)}$ \\
& {Not Mentioned}  & $118\ (46.1\%)$ & $24\ (3.0\%)$ & $\mathbf{142\ (13.4\%)}$ \\
\midrule
\multicolumn{5}{l}{Disease duration in years (range: $1 - 27$, mean: $7.39$, sd: $5.27$), n (\%)}\\
& \textless{}=2 & $14\ (5.5\%)$ \\
& $2-5$ & $42\ (16.4\%)$ \\
& $5-10$ & $36\ (14.1\%)$ \\
& $10-15$ & $19\ (7.4\%)$ \\
& $15-20$ & $8\ (3.2\%)$ \\
& \textgreater{}$20$ & $2\ (0.8\%)$ \\
& Unknown & $135\ (52.6\%)$ \\
\midrule
\multicolumn{5}{l}{Recording Environment, n (\%)} \\
 & \texttt{Home-Global} & $77\ (30.1\%)$ & $616\ (76.7\%)$ & $\mathbf{693\ (65.4\%)}$ \\
 & \texttt{PD Care Facility} & $118\ (46.1\%)$ & $24\ (3.0\%)$ & $\mathbf{142\ (13.4\%)}$ \\
 & \texttt{Clinic} & $47\ (18.3\%)$ & $28\ (3.5\%)$ & $\mathbf{75\ (7.1\%)}$ \\
 & \texttt{Home-BD} & $14\ (5.5\%)$ & $135\ (16.8\%)$ & $\mathbf{149\ (14.1\%)}$ \\
\bottomrule
\end{tabular}%
}}
\\[5mm] 
\subcaptionbox{\label{tab:test_demographics}}{
\resizebox{0.9\columnwidth}{!}{
\begin{tabular}{lcllll}
\toprule
\textbf{Cohort} && \textbf{Number of participants} & \begin{tabular}[c]{@{}l@{}}\textbf{With PD} \\ n (\%)\end{tabular} & \begin{tabular}[c]{@{}l@{}}\textbf{Age} \\ mean (range)\end{tabular} & \textbf{Female (\%)} \\ \hline
\begin{tabular}[c]{@{}l@{}}Cohort used in training \\ and cross-validation\end{tabular} & & $835$ & $195$ ($23.4\%$) & $61.5$ ($18 - 93$) & $56.2\%$ \\ 
\midrule
Clinic cohort for external testing && $75$ & $47$ ($62.7\%$) & $65.4$ ($18 - 86$) & $42.7\%$ \\ 
\midrule
Home-BD cohort for external testing && $149$ & $14$ ($9.4\%$) & $41.1$ ($18 - 80$) & $30.9\%$ \\ 
\bottomrule
\end{tabular}%
}}
\end{table}

\subsection*{Train-Test Splits}
In order to train, evaluate, and select the best machine learning model, we employed a two-step evaluation process. First, we determined the best-performing model using $k$-fold cross-validation on the \texttt{Home-Global} and \texttt{PD Care Facility} cohorts (the largest two cohorts in terms of participants with PD). Second, we assessed the model on the remaining two cohorts (i.e., \texttt{Clinic} and \texttt{Home-BD}) independently. These two settings had varied PD prevalence and socio-economic contexts, and the model has never seen data from these cohorts during training. The demographic information of the train and test cohorts is summarized in Table \ref{tab:test_demographics}.

\subsection*{Feature Extraction and Visualization}
We first extracted the facial action units~\cite{ekman1978facial} and facial landmarks from the videos using Openface~\cite{baltrusaitis2018openface} and Mediapipe~\cite{lugaresi2019mediapipe}. Both of these tools are open-source and widely used due to their efficiency, accuracy, and the ability to process video streams in real-time. The facial action units (AUs) are associated with the muscle movements of the face, and activation of a particular AU indicates the movement of a fixed set of facial muscles. For example, AU06, or ``Cheek Raiser'', involves raising the cheeks due to the contraction of the orbicularis oculi, pars orbitalis muscle around the eye socket. AU12, also known as ``Lip Corner Puller'', represents the movement caused by the zygomaticus major muscle, which pulls the corners of the lips upwards and outwards, creating a smile.
These two action units, when combined, are key indicators of a Duchenne smile~\cite{ekman1990duchenne}, a sincere and genuine smile associated with spontaneous joy and happiness. Since our dataset includes facial mimicry of disgusted and surprised expressions along with smiles, we measure multiple AU values (frame by frame for the videos).

While assessing signs of PD from a facial expression task, MDS-UPDRS instructs the clinician to observe the patient's \textit{eye-blinking frequency}, \textit{masked facies or loss of facial expression}, \textit{spontaneous smiling}, and the \textit{parting of lips at rest}. We engineered the facial features so that they reflect these four criteria. In our feature extraction setup, the expressiveness of each facial expressions is captured by four AUs. For example, AU01 (Inner Brow Raiser), AU06 (Cheek Raiser), AU12 (Lip Corner Puller), and AU14 (Dimpler) had three distinct peaks in smiling facial expression videos that ask participants to repeat the smile expression three times. Similarly, the expressiveness of a disgusted face is linked with AU04 (Brow Lowerer), AU07 (Eye Lid Tightener), AU09 (Nose Wrinkler), and AU10 (Upper Lip Raiser) and a surprised face is best represented with AU01 (Inner Brow Raiser), AU02 (Outer Brow Raiser), AU04 (Brow Lowerer), and AU05 (Upper Lip Raiser). In cases of masked facies, a characteristic of PD, we expect to see reduced intensity in these AUs. Although we don't capture resting face videos, our recordings include cycles of expressive and neutral expressions, allowing us to use the values of AU25 (Lips Part) and AU26 (Jaw Drop) to estimate a participant's ability to keep their lips together while their mouths are at rest. These two action units are consistently selected across all three facial expressions. Finally, AU45 (Blink) offers an objective measurement for the frequency of eye blinking during all three facial expression videos. This way, we extracted frame-by-frame values of seven facial action units for each expression videos. Finally, we used statistical aggregates --- mean, variance, and entropy --- of the AU intensity values to summarize them across the temporal dimension. 

In addition to the action unit features, we extracted facial attributes that simulate clinical assessments typically carried out in person as suggested in prior literature~\cite{gomez2021improving}. We used the face mesh solution of Mediapipe~\cite{lugaresi2019mediapipe} to extract $478$ 3D facial landmarks from a video frame and measured the ``opening of the left and right eye,'' ``rising of the left and right eyebrows," ``opening of the mouth,'' ``width of the mouth,'' and ``opening of the jaw.'' Since distances can vary depending on a participant's position in front the camera, these measures are normalized by the distance between two irises. Similar to AUs, we summarized these measures across all the video frames and calculated statistical aggregates. Combining the AUs and landmark features, we extracted 42 features for each facial expression, totalling to 126 features when all three expressions are used. Please refer to Supplementary Note 2 for more details.

To visualize the features for each of the three facial expressions (smile, disgust, and surprise) in our dataset, we first split the dataset based on these expressions and then applied Principal Component Analysis (PCA)~\cite{bro2014principal,adnan2021fast} to each subset. PCA is a dimensionality reduction technique that transforms the original features into a new set of uncorrelated variables called principal components (PCs), capturing the maximum variance in the data. For each expression, we identified the top two principal components and plotted them in a 2D scatter plot to visually assess whether the PD and Non-PD clusters could be separated. Additionally, we calculated the silhouette score~\cite{ROUSSEEUW198753} for each of the three facial expressions to quantify the quality of clustering. The silhouette score measures how similar an object is to its own cluster compared to other clusters, with a higher score indicating better-defined clusters. The application of PCA was essential as it reduces the complexity of high-dimensional data while preserving the most critical information, facilitating the identification of potential patterns or clusters. This approach enabled us to effectively visualize the data and highlighted that smile features hold a higher potential as a digital biomarker for PD.

\subsection*{Model Training and Evaluation}
In order to select the best model, we first perform an ablation study to evaluate different design choices using the $k$-fold cross-validation approach. 
We explored a variety of classifiers that have been proven effective in the literature for feature-based binary classification problems~\cite{mayr2014evolution,ferreira2012boosting,praveena2017literature,mendez2019comparative}, including Support Vector Machine (SVM), AdaBoost, Histogram-Based Gradient Boosting (HistBoost), XGBoost, and Random Forest.
We experiment with three different feature selection techniques from the literature -- logistic regression coefficients~\cite{cheng2006logistic,zakharov2011ensemble}, Boosted Recursive Feature Elimination (BoostRFE)~\cite{saberi2022lightgbm,jeon2020hybrid}, and  Boosted Recursive Feature Addition (BoostRFA)~\cite{hamed2018network}.
Since data scaling can boost the performance of machine learning models~\cite{ahsan2021effect,sharma2020biophysical}, we tried with two different feature scaling methods -- MinMax scaling and Standard scaling. 
Notably, we ensured that feature selection and scaling were performed based solely on the training data within each fold and not influenced by the test data.
Based on the ablation study, we found that ensemble~\cite{dong2020survey,dietterich2002ensemble,sagi2018ensemble,polikar2012ensemble} of multiple Histogram-Based Gradient Boosting models using smile features performed the best in both cross-validation and external test sets. 
Once we finalized the best-performing model, we proceed with the generalizabilty test using external test cohorts. 
Since the dataset is imbalanced, we employ SMOTE~\cite{chawla2002smote}, a widely-used approach to balance class distribution by generating synthetic minority samples. It is important to note that SMOTE was applied only to the training dataset. The test datasets, both in cross-validation and external testing, retained only the original samples to prevent any potential data leakage.
To evaluate model performance, we use a comprehensive set of metrics widely adopted by the interdisciplinary community of machine learning and healthcare: (i) Area Under the Receiver Operating Characteristic curve (AUROC), (ii) Accuracy, (iii) Sensitivity, (iv) Specificity, (v) Positive Predictive Value (PPV), and (vi) Negative Predictive Value (NPV). Additionally, we visualized the important features identified by the model using SHAP (SHapley Additive exPlanations)~\cite{NIPS2017_7062} analysis to provide insights into feature contributions and ensure model interpretability. SHAP values provide a unified measure of feature importance and offer insights into how each feature contributes to the model's predictions. Please see Supplementary Note 3 for further details regarding training, hyperparameter selection, and ablation experiments.

\subsection*{Statistical Analysis}
To estimate the confidence intervals (CIs) for our model's performance metrics, we employed a specialized bootstrapping approach. This involved running the model multiple times with the same best-performing hyperparameters but with 40 different random seeds. The random seed affects not only the data splits in cross-validation but also other aspects of the computation, such as histogram binning and the selection of features for each tree in the `HistBoost Classifier'. For each iteration, we recorded the performance metrics on the test set, and from these 40 bootstrap samples, we calculated the $95\%$ CIs for each metric. Specifically, the $95\%$ CIs were derived from the $2.5^{\text{th}}$ and $97.5^{\text{th}}$ percentiles of the bootstrapped distributions for each performance metric. This non-parametric method does not assume normality and provides a robust estimate of the CIs. Unless specified otherwise, all error bars shown and all intervals reported in the manuscript are $95\%$ CIs, and all significance testing were conducted at the $95\%$ confidence level. The metrics are reported in the format: mean $\pm$ standard error of the $95\%$ CI.

To determine whether the model exhibited different miss-classification rates across sex and ethnic subgroups, we employed a two-sample $Z$-test for proportions. This parametric test that assumes a sufficiently large sample size to approximate the normal distribution according to the central limit theorem (CLT)~\cite{kwak2017central}. Specifically, before proceed with the test, we ensured that the conditions $np \geq 5$ and $n(1-p) \geq 5$ were met for each sample, where $n$ is the sample size and $p$ is the proportion of interest, to ensure the validity of the test.

We also applied the two-sample Z-test to assess biases in underdiagnosis and overdiagnosis across sexes. However, during the underdiagnosis test, the sample size for non-white participants with PD was notably smaller, violating the Z-test assumptions. Therefore, we opted for Fisher's exact test\cite{upton1992fisher,bower2003use}, a non-parametric method well-suited for small sample sizes.
We also used the $Z$-test to assess underdiagnosis and overdiagnosis bias based on sex. 
However, during the test for underdiagnosis, the sample size for non-white participants with PD was significantly smaller, and the assumptions of the $Z$-test are not met, which might result in unreliable conclusion. Therefore, we opted for Fisher's exact test~\cite{upton1992fisher,bower2003use}, a non-parametric test that is well-suited for small sample sizes.
For analyzing the relationship between continuous variables, such as age or disease duration with model inaccuracy, we used Spearman's correlation test~\cite{rebekic2015pearson}, which is a non-parametric test to assesses the strength and direction of the association between two ranked variables. Unlike Pearson's correlation, Spearman's does not assume a linear relationship or normal distribution of the data, making it more suitable for the types of relationships we expected to observe in our dataset. 

When assessing the performance equity of our best-performing model across different demographic subgroups, we calculated confidence intervals for misclassification, underdiagnosis, and overdiagnosis rates using the normal approximation method, where prediction of each sample data is considered an independent Bernoulli random variable. 
Specifically, if $\hat{p}$ represents the metric of interest and $n$ is the sample size of the corresponding group, the $95\%$ C.I. is computed as $[(\hat{p} - 1.96 \times \sqrt{\frac{\hat{p}\times(1-\hat{p})}{n}}), (\hat{p} + 1.96 \times \sqrt{\frac{\hat{p}\times(1-\hat{p})}{n}})]$. Similar to the statistical significance test, in all cases, we ensured that the conditions of the CLT were satisfied before applying this method.



\section{Results}
\subsection*{Feature Selection} Out of the 126 features we computed for all three facial expression videos, 43 features were significantly different across participants with and without PD (at significance level, $\alpha = 0.01$), dominated by the features from the smile expression. Notably, mean and variance of ``AU12: lip corner puller'', mean of ``mouth width'', and variance of ``AU06: cheek raiser'' measured when a participant smiled had the most discriminating ability among the smile features, and the values were significantly different across the two groups (p-values: $10^{-24}, 10^{-7}, 10^{-8}, \text{and } 10^{-4}$, respectively). 

\subsection*{Identification of the Most Effective Facial Expression}

\begin{figure}[t]
    \centering
    \begin{subfigure}[b]{0.31\textwidth}
        \includegraphics[width=\textwidth]{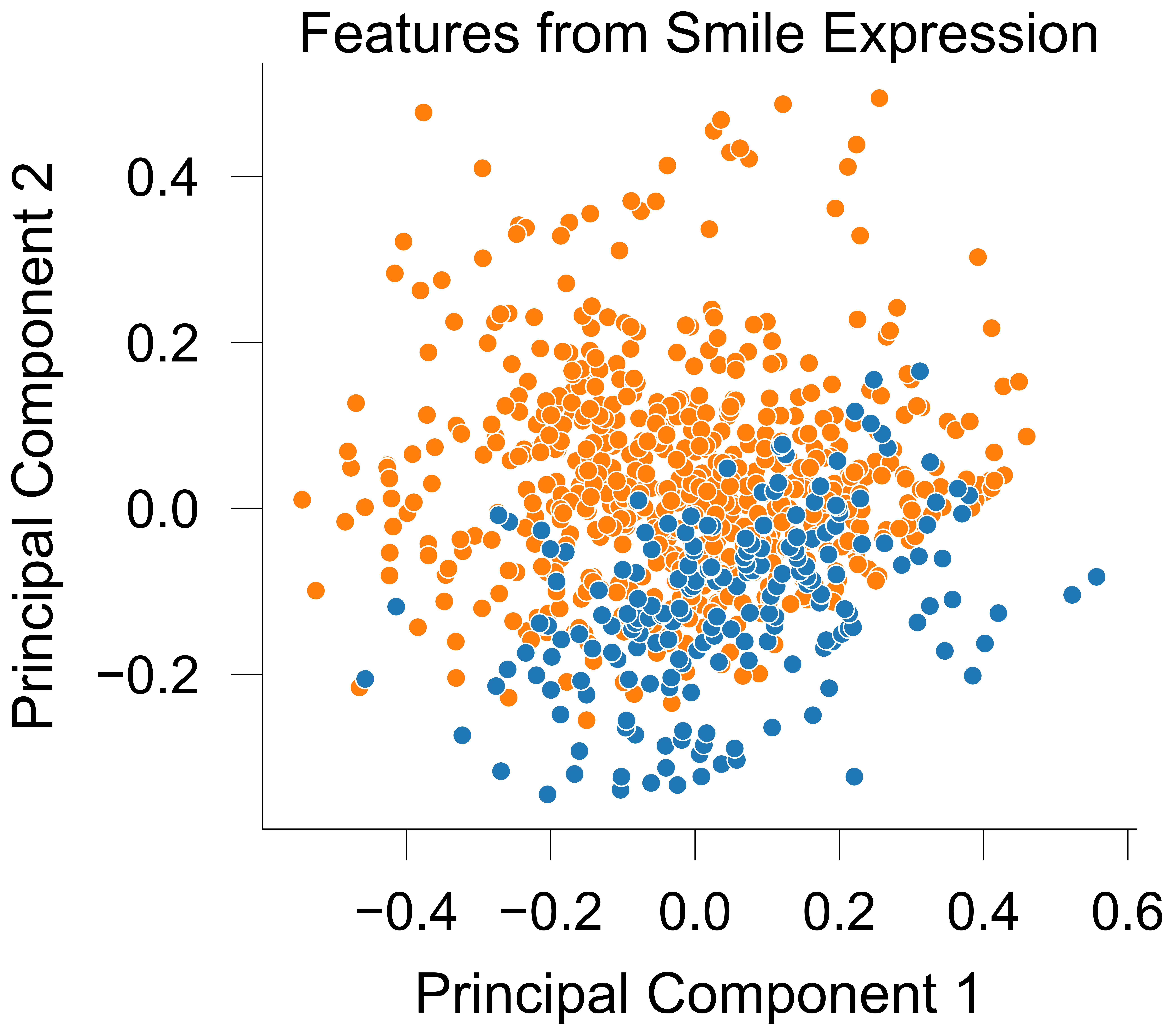}
        \caption{}
        \label{fig:smile}
    \end{subfigure}
    ~ 
    \begin{subfigure}[b]{0.31\textwidth}
        \includegraphics[width=.96\textwidth]{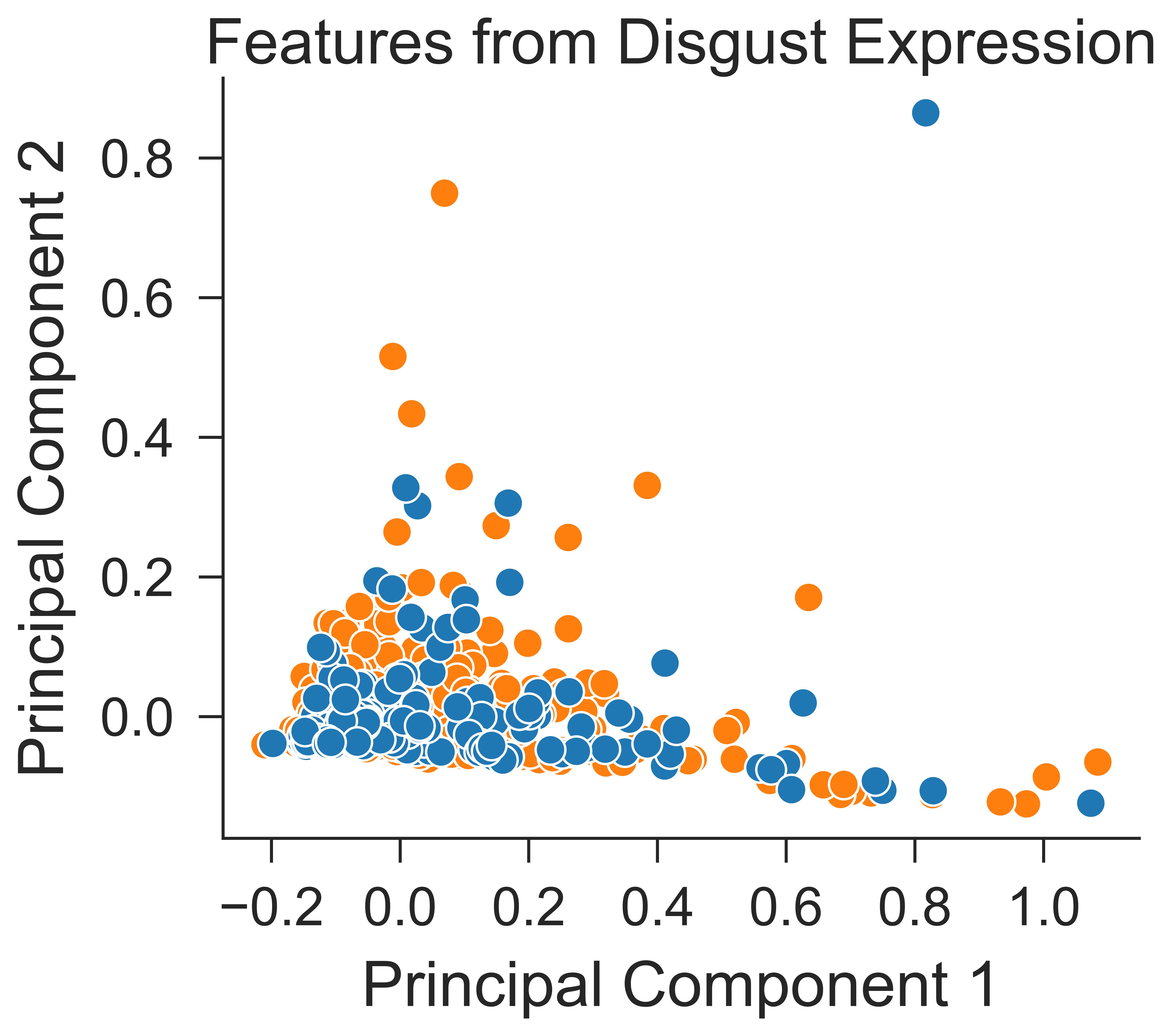}
        \caption{}
        \label{fig:disgust}
    \end{subfigure}
    ~ 
    \begin{subfigure}[b]{0.31\textwidth}
        \includegraphics[width=.99\textwidth]{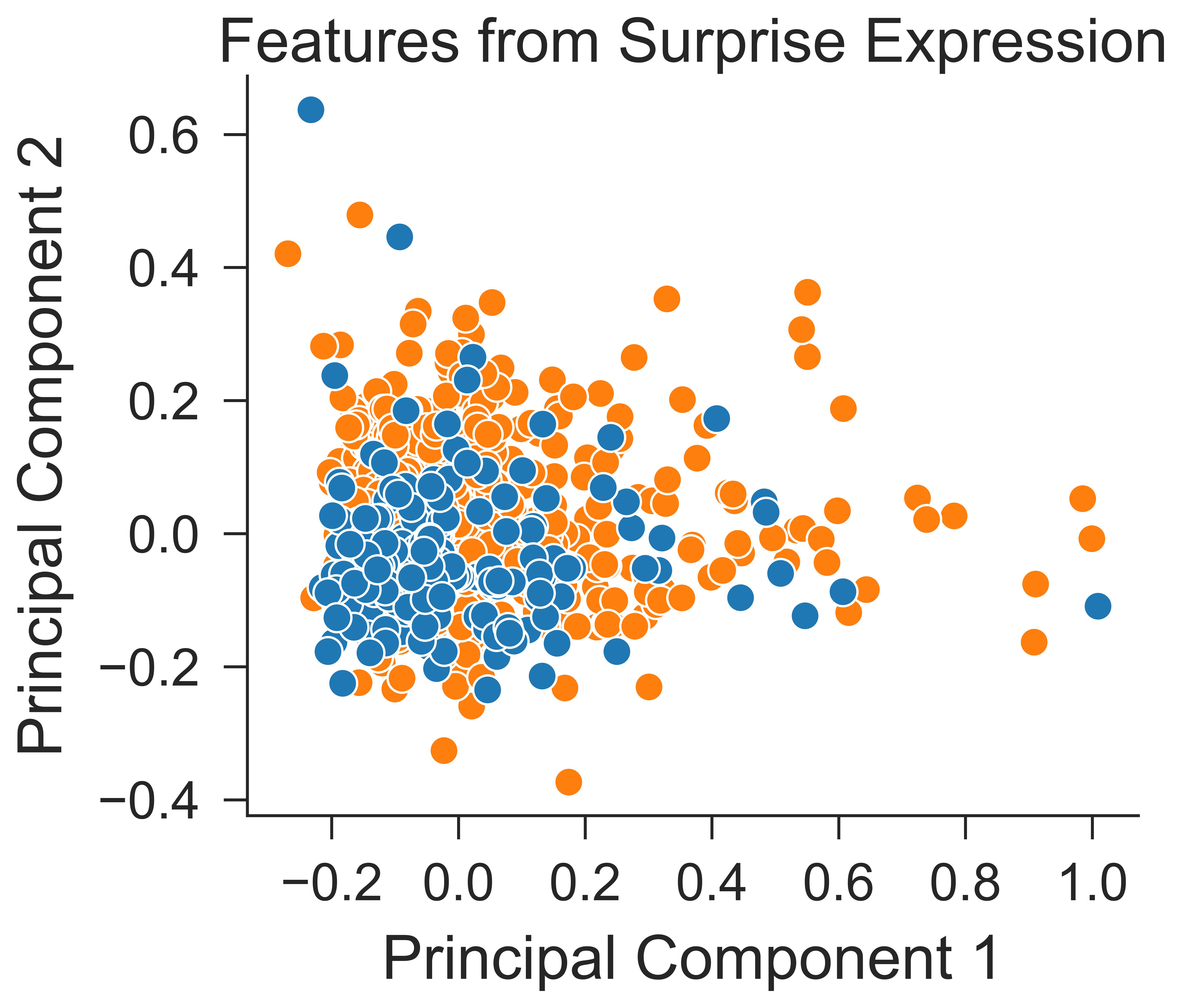}
        \caption{}
        \label{fig:surprise}
    \end{subfigure}
    ~
    
    \begin{subfigure}[b]{0.31\textwidth}
        \includegraphics[width=\textwidth]{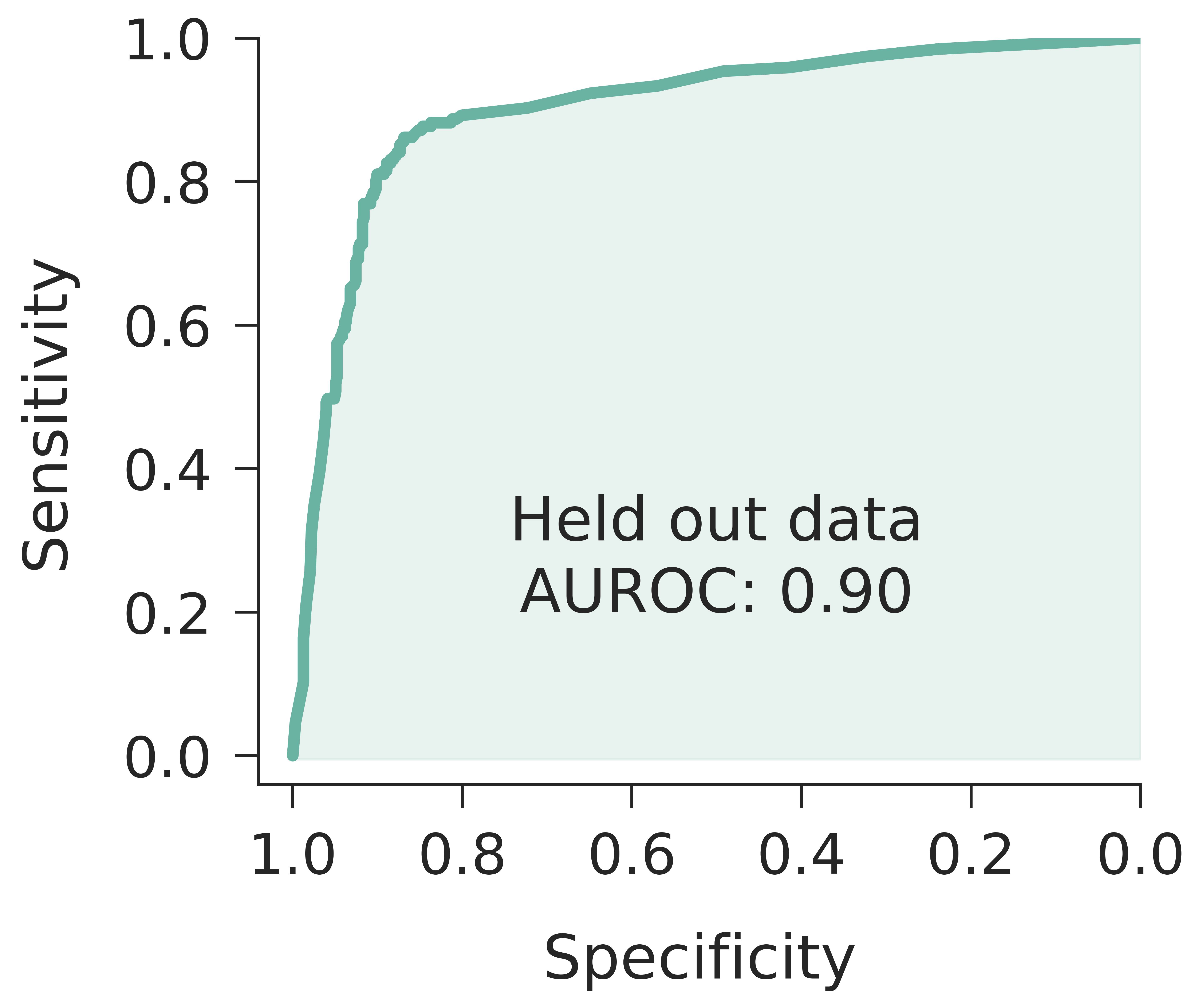}
        \caption{}
        \label{fig:roc_smile}
    \end{subfigure}
    ~ 
    \begin{subfigure}[b]{0.31\textwidth}
        \includegraphics[width=\textwidth]{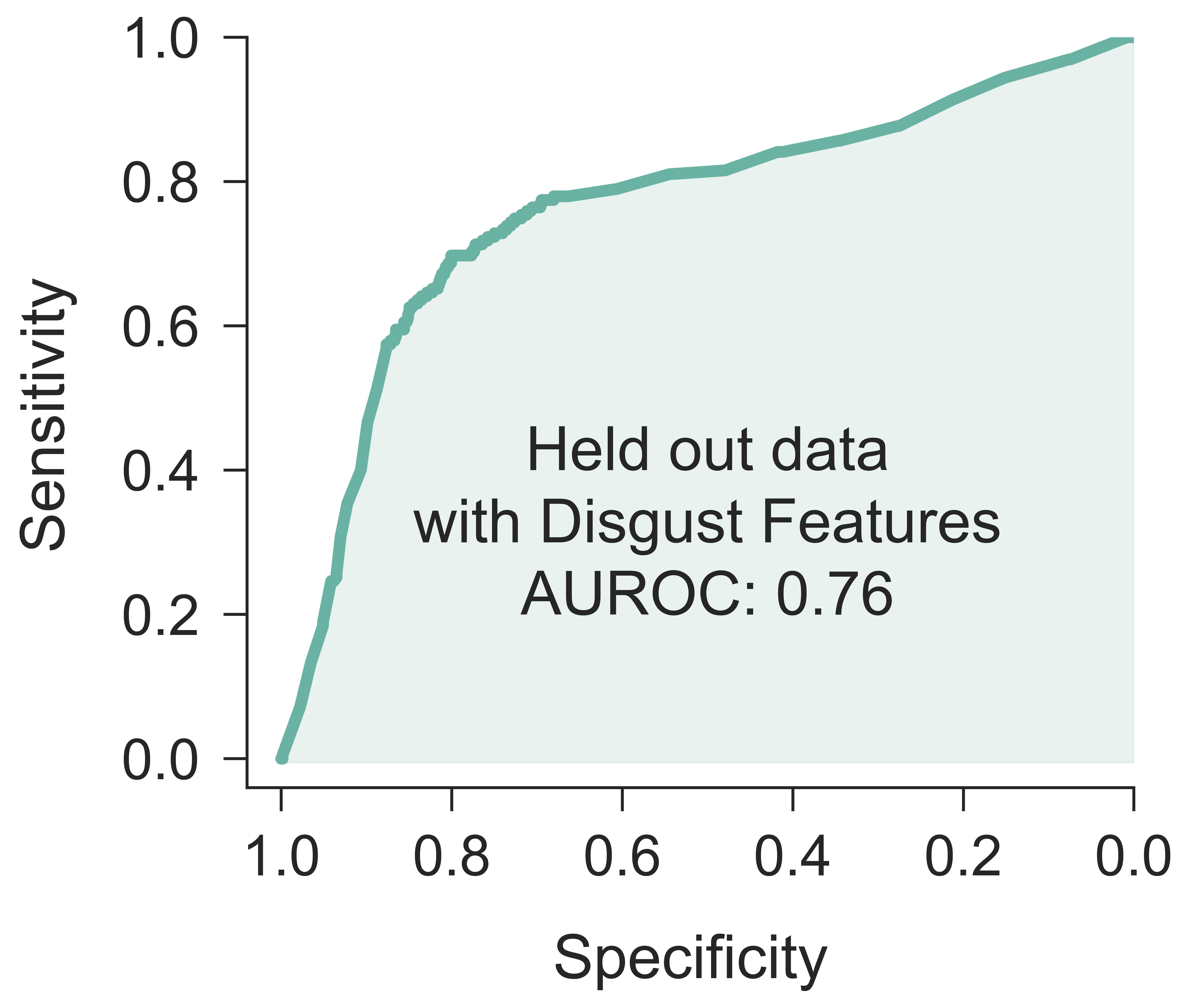}
        \caption{}
        \label{fig:roc_disgust}
    \end{subfigure}
    ~ 
    \begin{subfigure}[b]{0.31\textwidth}
        \includegraphics[width=\textwidth]{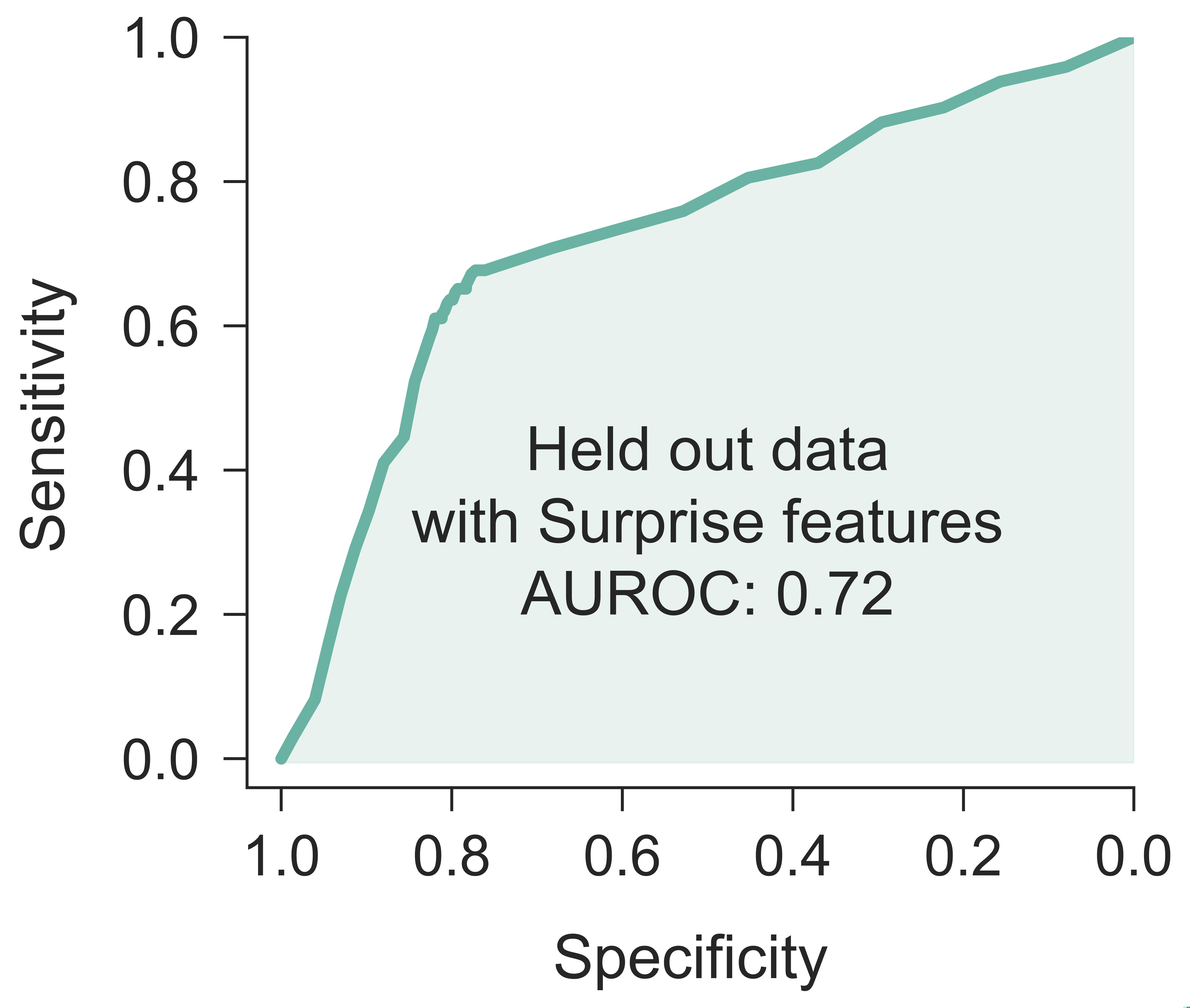}
        \caption{}
        \label{fig:roc_surprise}
    \end{subfigure}
    \caption{\textbf{Feature visualization and model performance for each facial expression.} All the extracted features from training videos ($n = 827$) of (a) smile, (b) disgust, and (c) surprise expression are reduced into two principal components for visualization. The orange and blue dots separate participants with and without PD. The ROC curve for the predictive model trained on features extracted only from the (d) smile, (e) disgust, and (f) surprise expression and evaluated on held-out data demonstrates how well the model can separate participants with and without PD using features from a single expression.}
    \label{fig:expressions}
\end{figure}

Our analysis into various facial expressions for PD yielded insightful findings. Visualization with PCA was utilized to assess the effectiveness of features from different expressions. The results revealed that digital features derived from smile expressions demonstrated the highest degree of separation between participants with and without PD, as evidenced by a silhouette score of \ita{0.18} (see Figure \ref{fig:expressions}). In contrast, features from disgust and surprise expressions displayed lower separability, with silhouette scores of \ita{0.11} and \ita{0.14}, respectively. These findings suggest that despite considering features from multiple expressions, the smile expression proved to be the most informative and reliable for PD identification.

\subsection*{Predictive Performance at Cross Validation} 
\begin{figure}[t]
  \begin{subfigure}[t]{.30\textwidth}
    \centering
    \includegraphics[width=1.0\linewidth]{roc_curve_Ensemble_with_LR.jpg}
    \caption{}
    \label{fig:roc_corve_lr}
  \end{subfigure}
  \hfill
  \begin{subfigure}[t]{.30\textwidth}
    \centering
    \includegraphics[width=1.0\linewidth]{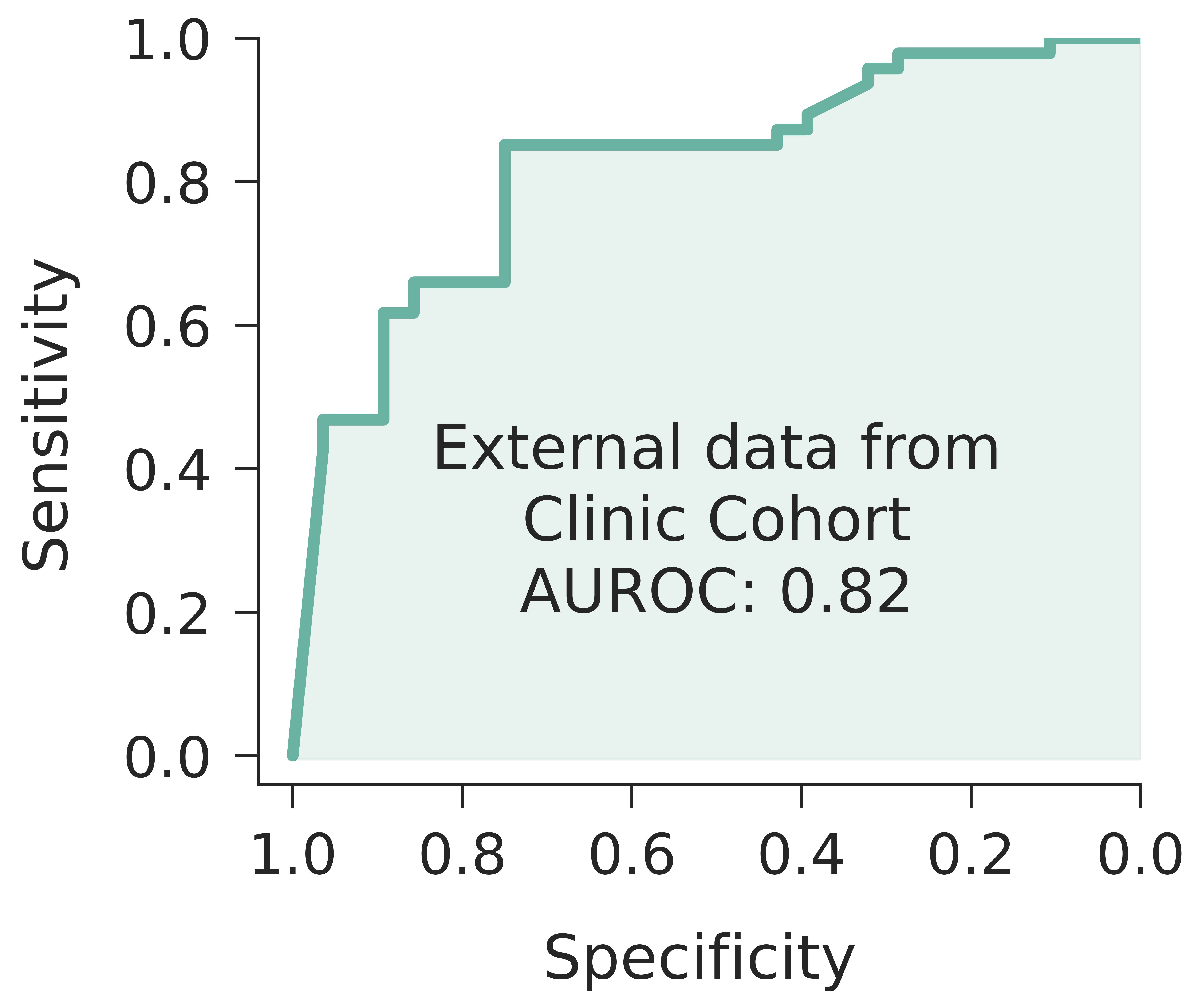}
    \caption{}
    \label{fig:roc_curve_clinic}
  \end{subfigure}
  \hfill
  \begin{subfigure}[t]{.30\textwidth}
    \centering
    \includegraphics[width=1.0\linewidth]{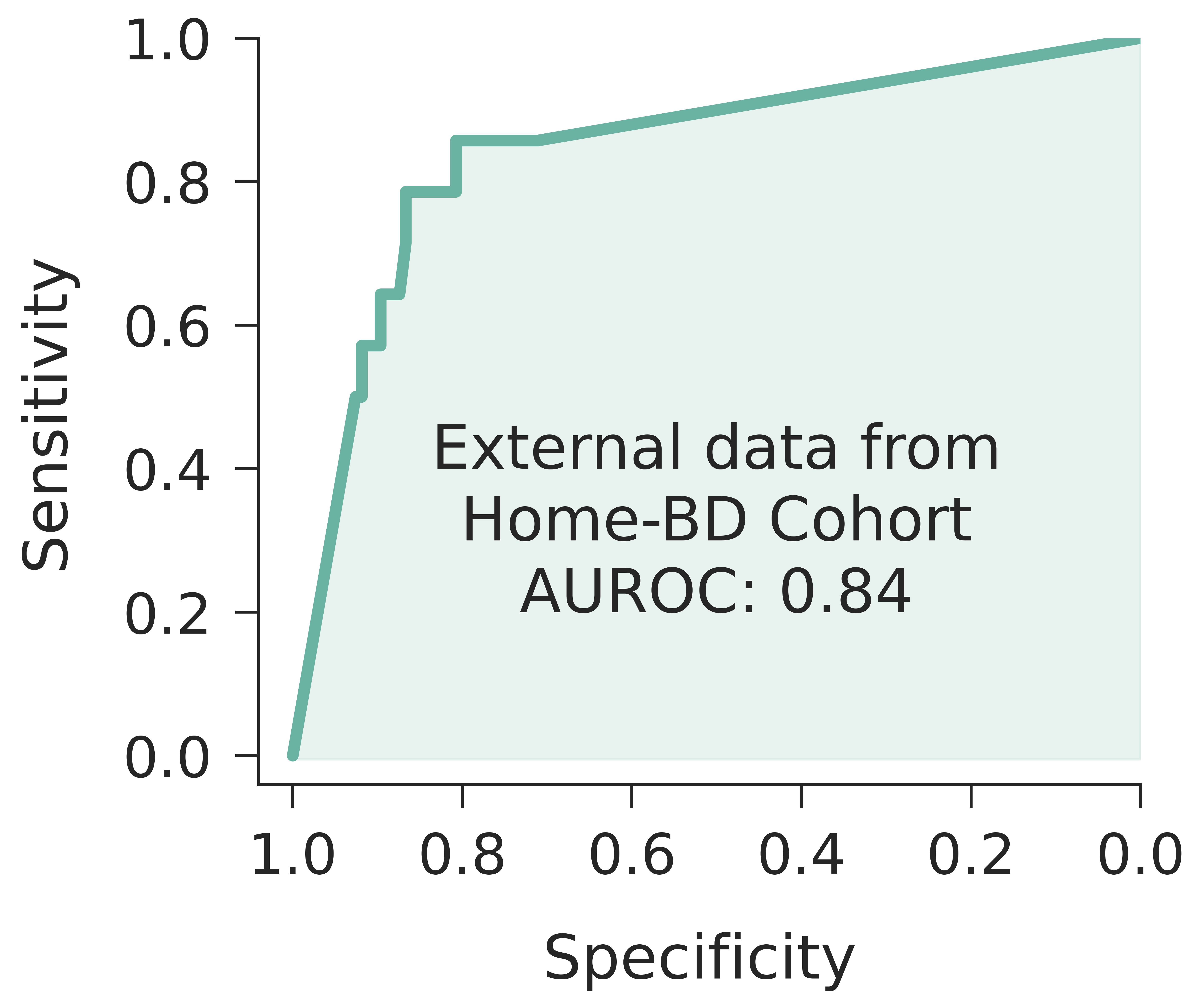}
    \caption{}
    \label{fig:roc_curve_home_bd}
  \end{subfigure}
  
  \bigskip
  
  \begin{subfigure}[t]{.30\textwidth}
    \centering
    \includegraphics[width=1.0\linewidth]{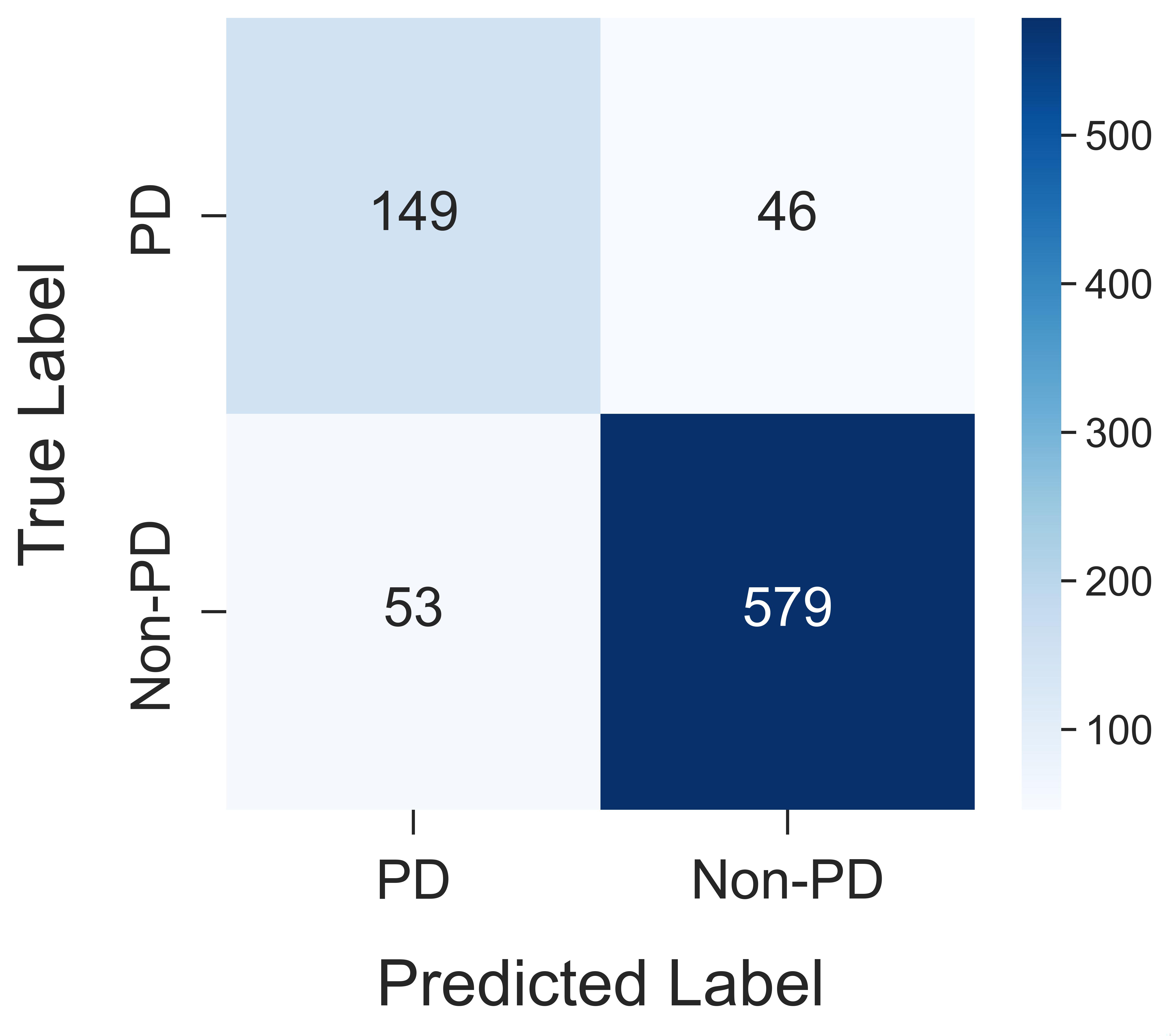}
    \caption{}
    \label{fig:cm_lr}
  \end{subfigure}
  \hfill
    \begin{subfigure}[t]{.30\textwidth}
    \centering
    \includegraphics[width=1.0\linewidth]{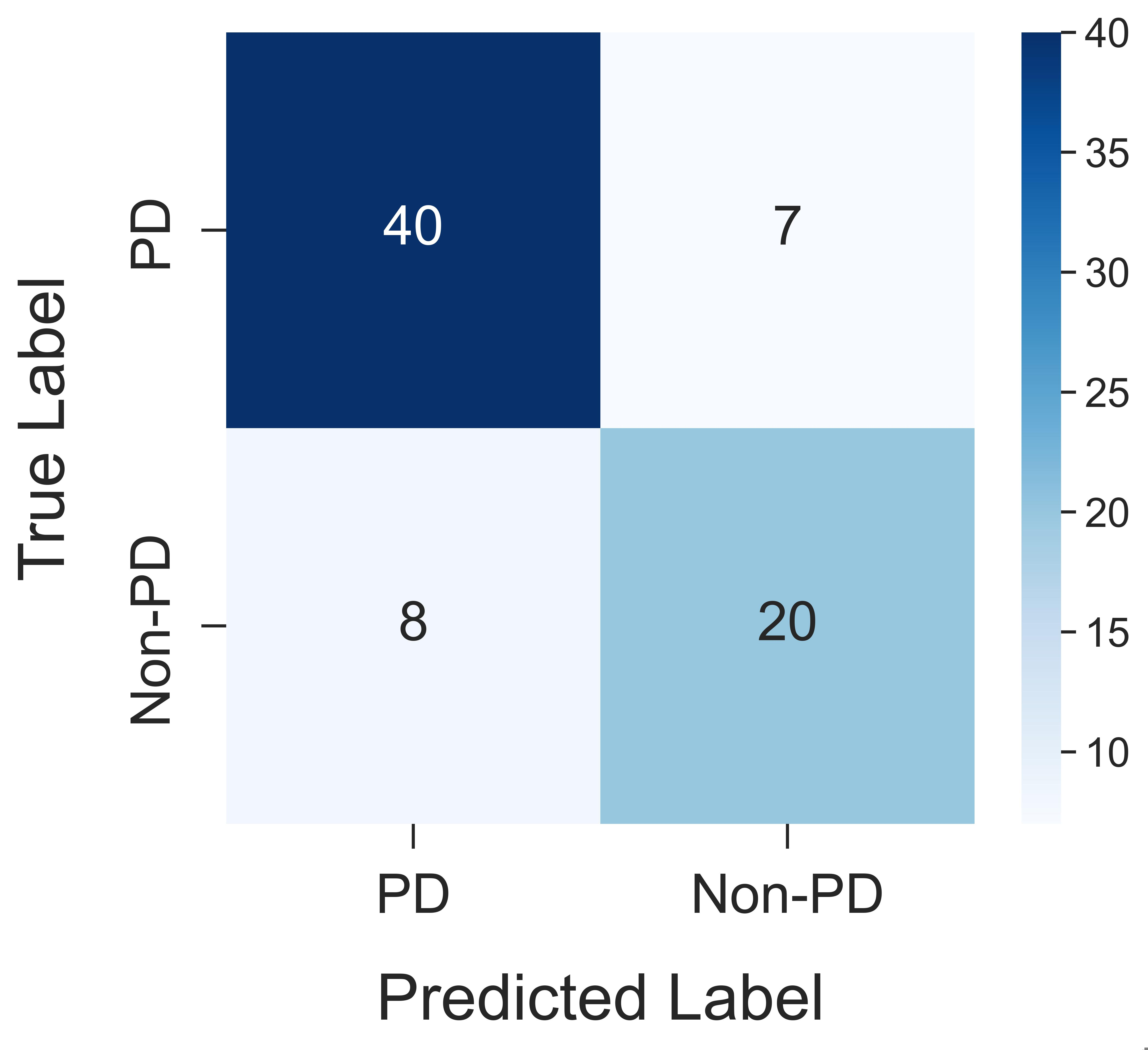}
    \caption{}
    \label{fig:cm_clinic}
  \end{subfigure}
  \hfill
  \begin{subfigure}[t]{.30\textwidth}
    \centering
    \includegraphics[width=1.0\linewidth]{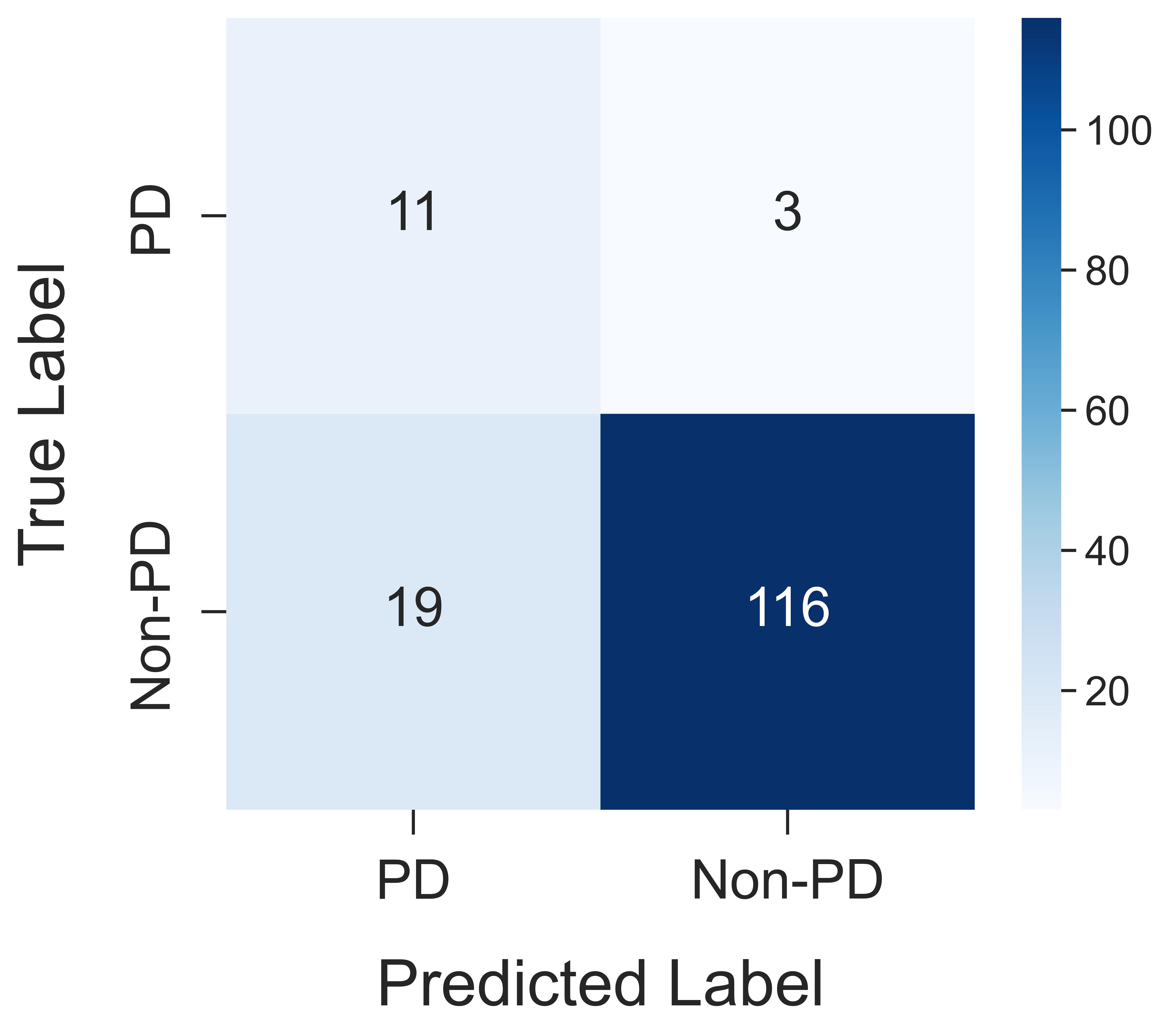}
    \caption{}
    \label{fig:cm_home_bd}
  \end{subfigure}

  \caption{\textbf{PD screening from videos of smiling.} ROC curves for differentiating between participants with
and without PD (a) on held-out data with k-fold cross validation ($n = 827$), (b) on external test data collected at a U.S. clinic
($n = 75$), and (c) on external home-recorded test data collected from Bangladesh ($n = 149$). The models only used features from the smile expression. The corresponding confusion matrices are displayed below the ROC curves (d-f).}
  \label{fig:performance}
\end{figure}

On held-out data ($10$-fold cross validation), the ensemble of 18 Histogram-Based Gradient Boosting models trained on smile features achieved an accuracy of \ita{$87.9 \pm 0.1\%$}, with an AUROC of \ita{$89.3 \pm 0.3\%$} (Figure \ref{fig:performance}). The model displayed a specificity of \ita{$91.4 \pm 0.3\%$} and a sensitivity of \ita{$76.8 \pm 0.4\%$}. Additionally, the PPV of \ita{$73.3 \pm 0.5\%$} and the NPV of \ita{$92.7 \pm 0.1\%$} further reinforce the model's reliability in predictive performance.


\subsection*{Model Generalization} For generalizability testing, we evaluated the ensemble model on the \texttt{Clinic} and \texttt{Home-BD} cohorts not included in the training data. The model achieved an accuracy of \ita{$79.6 \pm 1.1\%$} and an AUROC score of \ita{$81.7 \pm 0.9\%$} when evaluated on the \texttt{Clinic} cohort. The sensitivity, specificity, PPV, and NPV on this external test data were \ita{$84.9 \pm 0.6\%$}, \ita{$71.2 \pm 1.5\%$}, \ita{$83.1 \pm 0.9\%$}, and \ita{$74.0 \pm 2.3\%$} respectively. For the external \texttt{Home-BD} test set, the model achieved an accuracy of \ita{$84.8 \pm 0.5\%$} and an AUROC score of \ita{$81.2 \pm 1.0\%$}. \ita{Note that while the specificity ($86.3 \pm 0.3\%$), sensitivity ($71.1 \pm 1.5\%$), and NPV ($96.8 \pm 0.1\%$) remained competitive, we observed a sharp decline in PPV ($35.6 \pm 3.5\%$ on the \texttt{Home-BD} setting compared to $73.3 \pm 0.5\%$ on held-out data).}

\begin{figure}[t]
  \centering
  \begin{subfigure}[t]{.48\textwidth}
    \centering
    \includegraphics[width=1.0\linewidth]{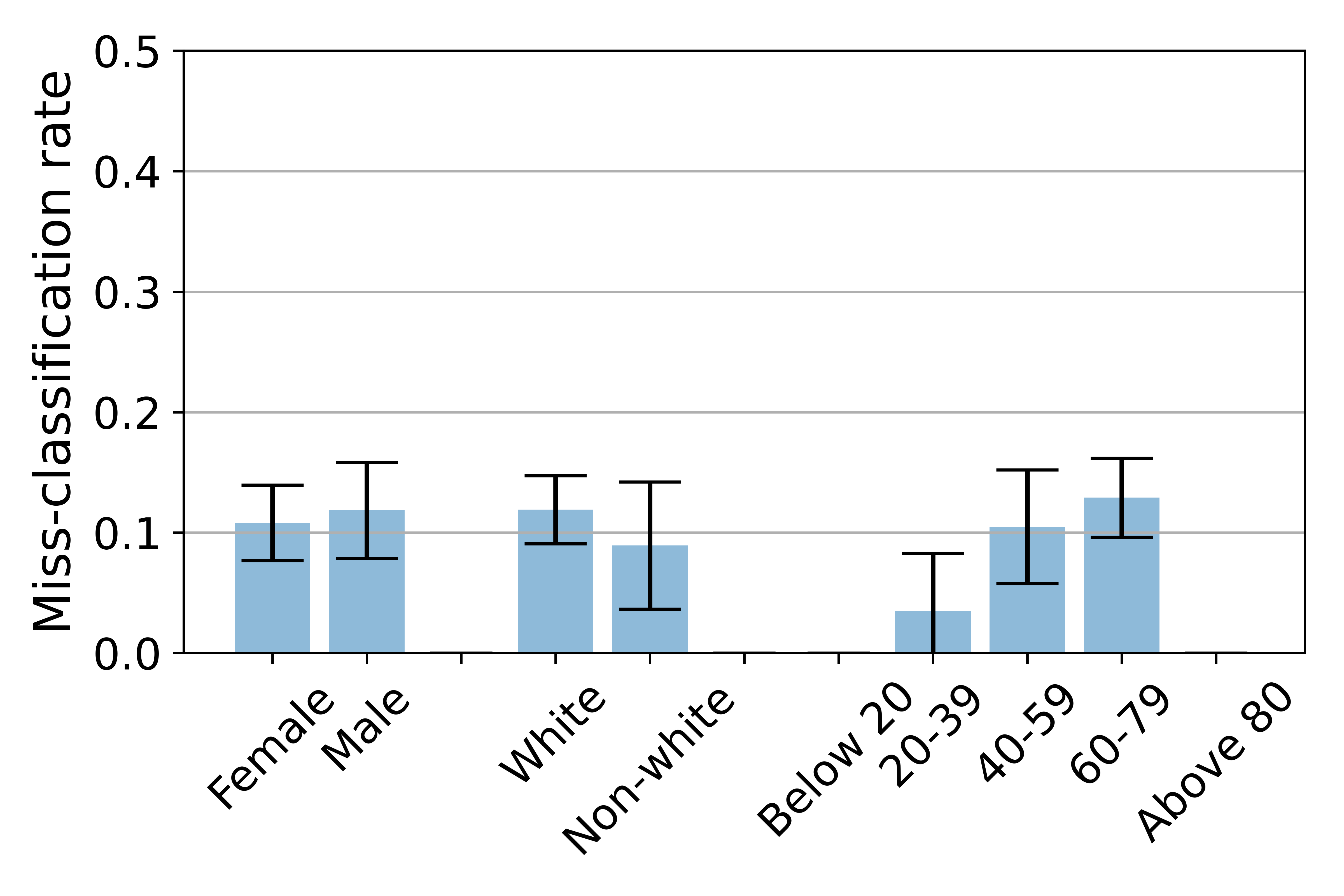}
    \caption{}
    \label{fig:missclassification_bias}
  \end{subfigure}
  \hfill
  \begin{subfigure}[t]{.48\textwidth}
    \centering
    \includegraphics[width=1.0\linewidth]{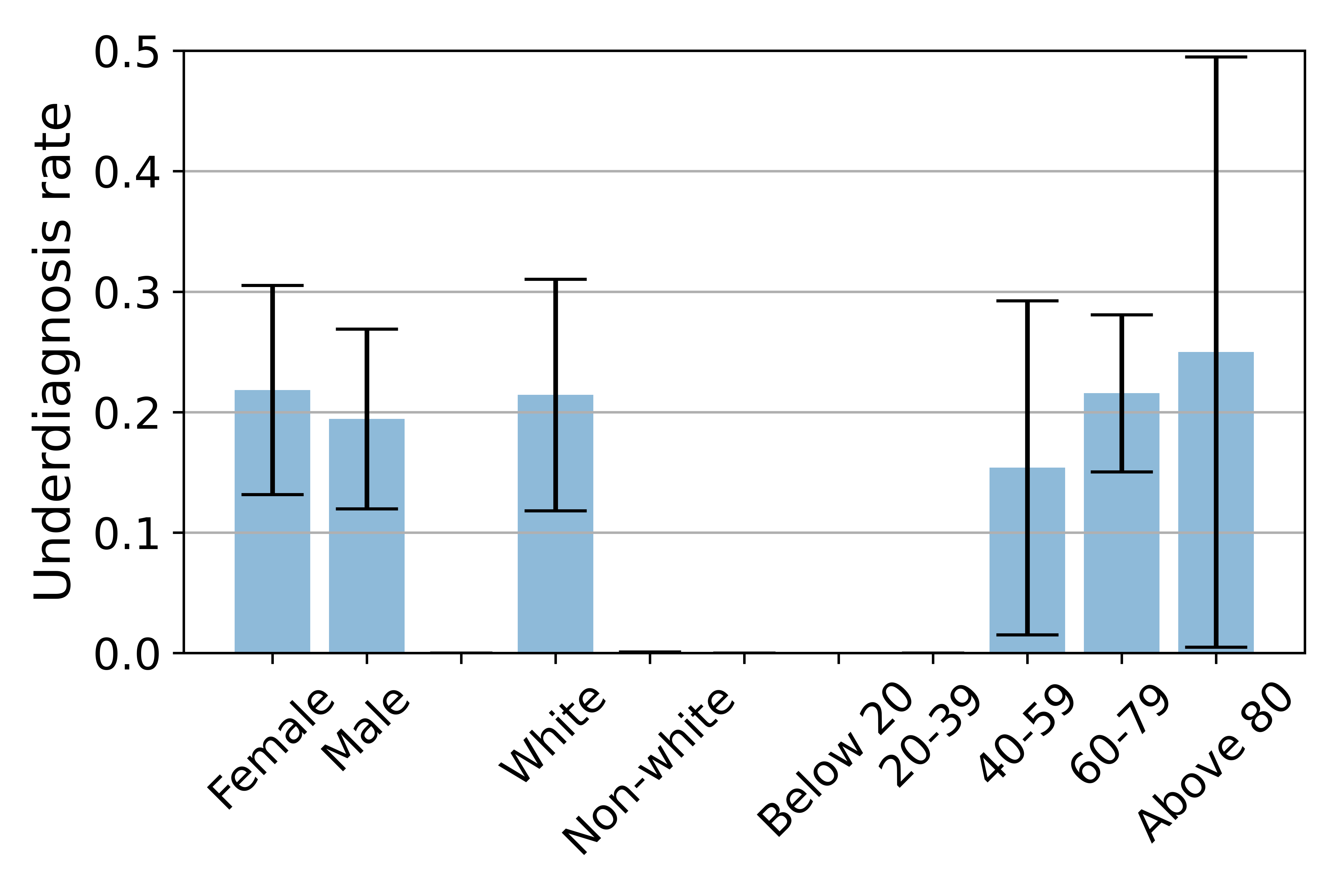}
    \caption{}
    \label{fig:underdiagnosis_bias}
  \end{subfigure}
  
  \begin{subfigure}[t]{.48\textwidth}
    \centering
    \includegraphics[width=1.0\linewidth]{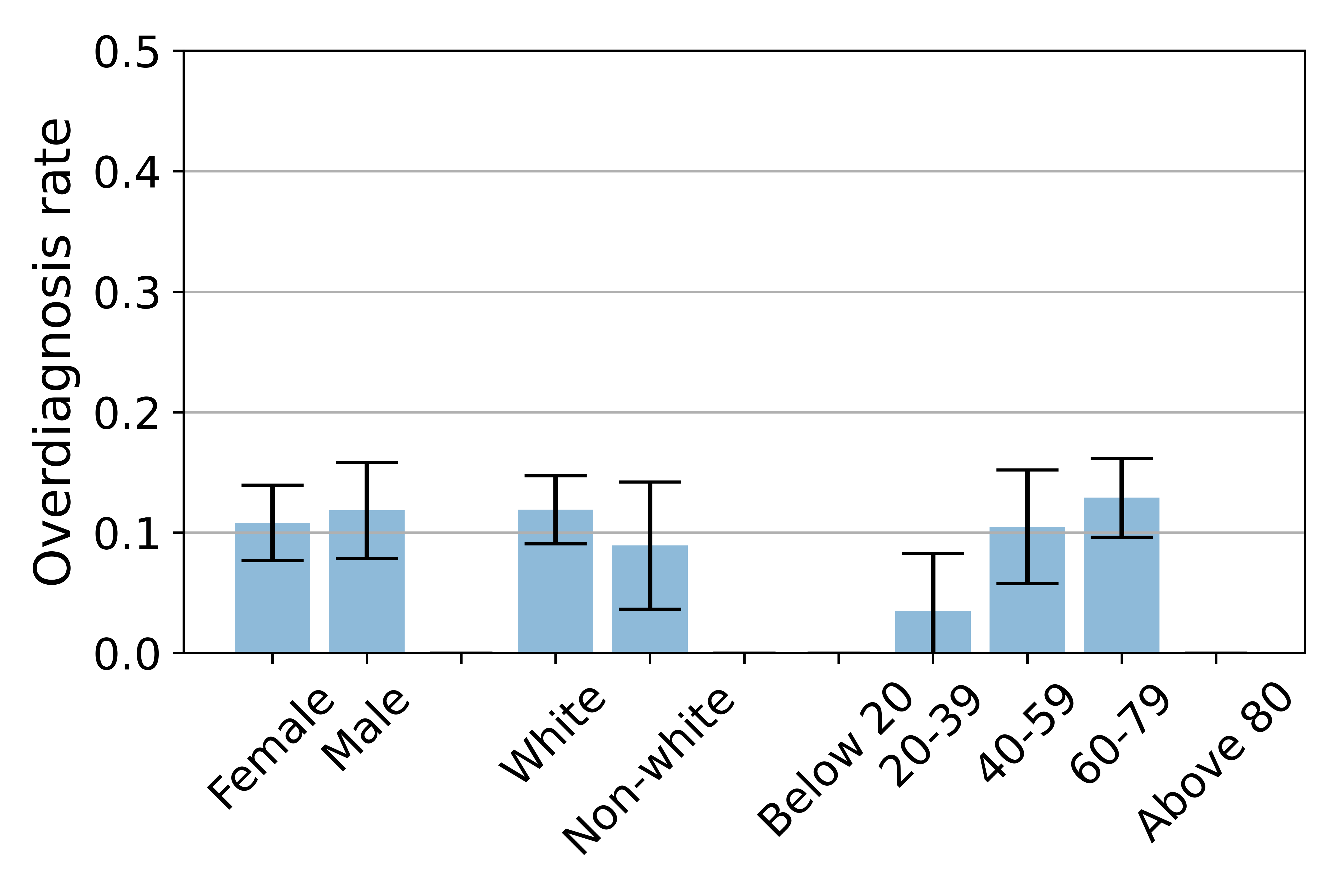}
    \caption{}
    \label{fig:overdiagnosis_bias}
  \end{subfigure}
  \hfill
  \begin{subfigure}[t]{.48\textwidth}
    \centering
    \includegraphics[width=1.0\linewidth]{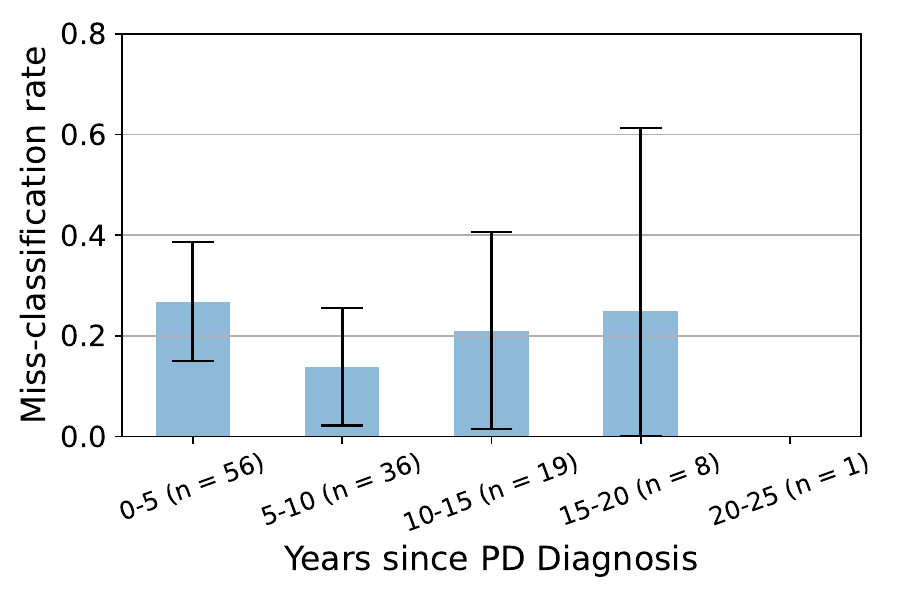}
    \caption{}
    \label{fig:error_vs_duration}
  \end{subfigure}
  
  \caption{\textbf{Comprehensive analysis of model performance across subgroups and PD durations.} The first three plots demonstrate the rate of (a) miss-classification, (b) underdiagnosis, and (c) overdiagnosis across population subgroups based on sex, ethnicity, and age. The last plot visualizes the (d) miss-classification rate across different bins of disease duration. Durations are binned into yearly intervals, with sample sizes for each interval noted in parentheses. In all cases, The error bars represent $95\%$ confidence interval.}
  \label{fig:comprehensive_analysis}
\end{figure}

\subsection*{Statistical Bias Analysis} 
\textbf{On Held-out Data.} The predictive model exhibited a miss-classification rate of $14.1 \pm 3.6\%$ for male participants ($n = 361$) and $12.9 \pm 3.0\%$ for female participants ($n = 466$). The difference was not statistically significant (test statistic, $z = 0.52$, $\text{p-value} = 0.60$). 
Similarly, there was no detectable bias when comparing white participants ($n = 574$) with Non-whites ($n = 119$), as the average miss-classification rates were $13.1 \pm 2.8\%$ and $8.4 \pm 5.0\%$, respectively ($z = 1.41$, $\text{p-value} = 0.16$). 
Age was found to be positively correlated with the miss-classification rate (Spearman's rank correlation, $\rho = 0.35$, $\text{p-value} = 0.004$) meaning the model demonstrated lower accuracy for older participants, {and it did not occur by chance}. 
We did not observe any underdiagnosis bias of the predictive model at a statistically significant level based on sex and ethnicity. Specifically, the underdiagnosis rates for the male ($n = 108$) and female ($n = 87$) participants were $19.4 \pm 7.5\%$ and $21.8 \pm 8.7\%$, respectively, showing no significant difference ($z = 0.41$, $\text{p-value} = 0.68$). 
Again, $21.4 \pm 9.6\%$ ($n = 70$) white participants were underdiagnosed by the model, while the rate was surprisingly $0.0\%$ ($n = 7$) for non-white participants. Based on Fisher's exact test, this difference was also insignificant (Fisher's odd ratio $ = 0.0$, $\text{p-value} = 0.33$). 
{We identified a slight positive correlation ($\rho=0.28$) between age and underdiagnosis rate. However, this correlation was not statistically significant ($\text{p-value} = 0.08$), suggesting the observed relationship could be due to random chance. Therefore, we cannot conclude that age is related to the underdiagnosis rate based on this result.
Again, we did not detect any overdiagnosis bias based on sex ($253$ male participants, $379$ female participants, $z\text{-score} = 0.41$, $\text{p-value} = 0.69$) and ethnicity ($504$ white participants, $112$ non-white participants, Fisher's odd ratio $= 0.90$, $\text{p-value} = 0.37$).
The older participants were slightly more overdiagnosed by the predictive model as we observed a modest positive correlation between age and overdiagnosis rate ($\rho = 0.23$). However, with a non-significant $\text{p-value}$ of $0.07$, we cannot assert a definitive connection between age and the overdiagnosis rate based on this result.
Figure \ref{fig:missclassification_bias}, \ref{fig:underdiagnosis_bias}, and \ref{fig:overdiagnosis_bias} provide a visual overview of the subgroup-based analyses.

\textbf{On External Validation.} When evaluated on the \texttt{Clinic} cohort, the model exhibited a miss-classification rate of $25.0 \pm 15.0\%$ ($n = 32$) for female participants, compared to only $16.3 \pm 11.0\%$ for males ($n = 43$). But this difference was not statistically significant according to a two-sample $Z$-test (for proportions) ($z = 0.93$, $\text{p-value} = 0.35$).  
On the \texttt{Home-BD} cohort, the model miss-classified male participants ($n = 103$) $19.4 \pm 7.6\%$ times while the rate was only $4.3 \pm 5.9\%$ for the females ($n = 46$). This was statistically significant based on two-sampled $Z$-test ($Z = 2.40$ and a p-value of $0.01$), indicating male participants were significantly more miss-classified compared to females. 
For both settings, the miss-classification rate was not significantly correlated with age
($\rho = -0.32 \text{ and } 0.10$ for \texttt{Clinic} and \texttt{Home-BD} settings, respectively, with corresponding p-values of $0.07$ and $0.87$). Supplementary Figure 1(a) and 1(b) provide a visual overview of the subgroup-based analyses on the external test datasets.

\textbf{Performance Across Disease Duration.} In addition to conducting group-based statistical analyses, we sought to ensure that our model performs equitably for participants with PD at various stages of the disease. Although specific information on PD stages was not collected in this study, we obtained disease duration data for $121$ individuals with PD ($45$ from the \texttt{Home-Global} cohort and $76$ from the \texttt{Clinic} cohort) out of a total of $256$ participants. For the analysis based on disease duration, we combined results from both the held-out \texttt{Home-Global} cohort and the external \texttt{Clinic} cohort.
We calculated the Spearman's correlation coefficient between the duration of the disease and the average miss-classification rate. The analysis revealed a moderate negative correlation coefficient ($\rho = -0.35$), suggesting that error rates tend to decrease as the disease duration increases. However, the associated $p$-value of $0.15$ indicates that this correlation is not statistically significant. Therefore, we cannot confidently assert a real association between disease duration and error rate based on this data. Furthermore, we categorized the $121$ participants into five duration-based groups and performed a Chi-Square test to determine if there are significant differences in miss-classification rates across these groups. The test resulted in a $\chi^2$ value of $2.89$ with a $p$-value of $0.64$, indicating that the observed differences in error rates across the groups are not statistically significant. This suggests that any variations in error rates are likely due to random chance rather than true differences between the groups. Figure~\ref{fig:error_vs_duration} demonstrate the average miss-classification rate by the predictive model across different disease duration groups.

\subsection*{Feature Importance Visualization} 
To visualize the importance of various features in our dataset to the model's prediction, we generated the SHAP beeswarm plot, shown in Figure 4. This plot illustrates the influence of the top features, such as AU12 mean, AU12 entropy, and mouth-width mean, on the model's predictions. Additionally, the ``Sum of 15 other features'' combines the contributions of less significant features, providing a comprehensive overview of the model's decision-making process. In the figure, each dot represents a SHAP value for a specific feature across all samples. The horizontal position of each dot indicates the SHAP value, reflecting the feature's impact on the model's prediction. The color gradient of the dots, ranging from blue to red, represents the feature values, with blue indicating lower values and red indicating higher values. For instance, we see that lower values of ``AU12 mean'' have positive SHAP values (the points extending towards the right are increasingly blue) and higher values of it have negative SHAP values (the points extending towards the left are increasingly red). This indicates that lower AU12 mean values lead to higher chances of PD prediction and vice versa. Similarly, for the feature ``mouth-open mean,'' higher values have positive SHAP values (red dots extending to the right), indicating that higher mouth-width mean values are associated with a higher likelihood of predicting PD.

\begin{figure}[t]
    \centering
    \includegraphics[width=0.85\textwidth]{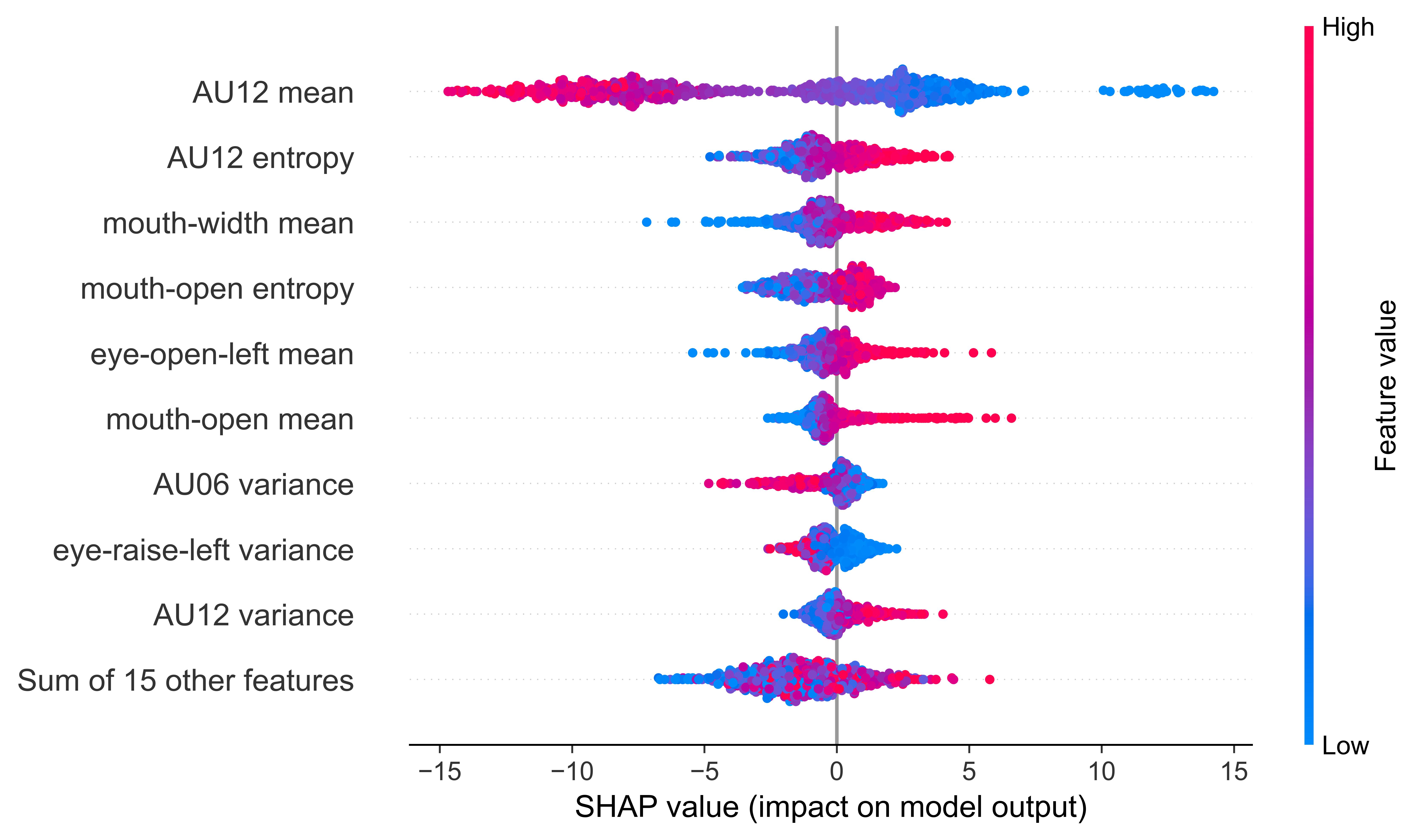}
    \caption{SHAP beeswarm plot showing the impact of features on the model's output. The plot displays the top features contributing to the model, with the color representing the feature value (red for high, blue for low). The "Sum of 15 other features" combines the contributions of less significant features.}
    \label{fig:shap_beeswarm}
\end{figure}

\subsection*{Predictive Performance with Three Expressions} Finally, in addition to the default best model that only learns from the smile videos, we also experimented with models trained on surprise and disgust expression features (and all subsets of three expressions). However, the performance of these models was significantly worse than the models trained on smile expression, which aligns with our PCA based visualization.
Among all other subsets of these three expressions, concatenating features of all expressions performed favorably on held-out data
with an accuracy of $89.5 \pm 0.1\%$, and an AUROC of $89.1 \pm 0.2\%$. 
However, when tested on the data from the \texttt{Clinic} cohort, 
the accuracy dropped to $71.7 \pm 0.8\%$ (an absolute decrease of $17.8\%$) and the AUROC score dropped to $72.4 \pm 1.0\%$ (an absolute decrease of $17.1\%$). Similarly, when tested on data collected from Bangladesh, the predictive model achieved an accuracy of $79.2 \pm 0.4\%$ and an AUROC of $76.9 \pm 0.9\%$. 
Please see Supplementary Table 2 where we report model performance for all the design choices.

\section{Discussion}
Currently, diagnosing PD necessitates clinical visits, which are not only time-intensive and can be expensive but frequently pose challenges in terms of accessibility. These limitations may deter individuals from seeking medical attention until their PD symptoms have considerably advanced. New, more accessible ways of detecting PD are needed to support earlier diagnosis and ultimately improve the care of those living with PD. Here, we found that AI-models trained on recordings of smile features can accurately detect PD. Given the ubiquity of smartphones capable of capturing video recordings, this method holds promise for supporting the diagnosis of PD. Currently, the number of smartphones all around the world is approximately 7.2 billions~\cite{taylor2023number}, which is expected to exceed 7.7 billions by the year 2028~\cite{taylor2023smartphone}. It is possible to imagine a mobile application that, with user consent, autonomously, passively assesses the intricacies of their facial expressions and facilitates a referral for clinical evaluation when needed. Our study represents one step towards a model that provides accessible, cost-effective, early PD screening.

The generalizability of models trained on smile features alone is an interesting finding. However, it supports the well-known Occam's razor~\cite{blumer1987occam} that simplicity favors generalizability. Although adding features from expressions enables the predictive model with more information, this adds to the complexity of the model. Unless the additional features are highly informative, it is likely that they will push the model towards overfitting -- performing better on training data or data with similar distribution and substantially worse on data with a distribution shift. Indeed, with the concatenation of the features from all three expressions, we observe a similar scenario -- the model achieved $89.5 \pm 0.1 \%$ accuracy and $89.1 \pm 0.2 \%$ AUROC on held-out data, while the performance on external cohorts dropped as much as $17.8\%$. 
The feature visualization with PCA, further illustrated that the features extracted from the smile task are relatively more separable between participants with and without PD compared to the disgust.
This may  explain why the model trained on smile expression features performed better than the other expressions. 


The SHAP analysis identified the top key features that correspond closely with the MDS-UPDRS criteria for assessing PD. For facial expressivity, features such as AU12 mean, AU12 entropy, and AU06 variance emerged as highly significant. The AU12 feature, which captures the activation of the lip corner puller muscles, is essential for expressing emotions like smiling. Spontaneous smile is strongly associated with AU12 mean and AU12 entropy, which quantify the intensity and variability of muscle activation involved in smiling. The ability to keep the mouth closed at resting time is approximated by features such as mouth-width mean and mouth-open mean, which measure the average width of the mouth and its openness, respectively, thereby evaluating lip closure ability. Eye blinking is reflected in features such as eye-open-left mean and eye-raise-left variance; the mean eye-opening feature indicates the typical state of eye openness, while the variance in eye-raising provides insights into the variability of eye movements. These findings demonstrate that our model effectively leverages relevant physiological signals to generate predictions that align with established clinical criteria. Additionally, the impact of these features aligns with their clinical relevance in indicating PD. For instance, lower AU12 mean values indicate less expressiveness of the smile, which contributes more significantly to PD prediction. The SHAP analysis also highlighted the feature mouth-open mean, where higher values suggest a reduced ability of PD individuals to keep their lips closed during resting times, thereby increasing the likelihood of PD prediction. This concordance between digital feature importance and clinical understanding underscores the robustness and interpretability of our model in predicting PD.

The performance of our models reported in different settings is comparable to clinicians~\cite{rizzo2016accuracy}. When PD diagnosis is performed by non-expert clinicians (i.e., general practitioners, neurologists, and geriatricians), the average accuracy is $73.8\%$ ($95\%$ credible interval (CrI) $67.8\% - 79.6\%$). The average accuracy of movement disorder specialists is $79.6\%$ ($95\%$ CrI $46\% - 95.1\%$) at initial assessment and $83.9\%$ ($95\%$ CrI $69.7\% - 92.6\%$) after follow-ups. The accuracy of our model relying on features from the smile expression ranges from $79.6 \pm 1.1\%$ to $87.9 \pm 0.1\%$ depending on experiment setups. Furthermore, the benefit of this study is not limited to PD alone. This research suggests that  technology could be used in the diagnosis of facial muscle-related diseases such as blepharospasm~\cite{fan2022case,tolosa2006dystonia}, Meige syndrome~\cite{weiner1982meige}, and Bell's palsy~\cite{savica2009bell,deng2007family}. By allowing individuals to perform initial assessments at home, our proposed invention has the potential to expedite the assessment of these disorders as well. 

When tested on the external test dataset (completely unseen during the training process), the model demonstrated a notable performance drop.
Compared to an accuracy of $87.9 \pm 0.1\%$ on held-out data, the accuracy in \texttt{Clinic} and \texttt{Home-BD} settings were $79.6 \pm 1.1\%$, and $84.8 \pm 0.5\%$, respectively.
It is encouraging to note that these results are promising as they fall within the established $95\%$ credible intervals by the clinicians.
However, when tested on the cohort from Bangladesh, a significant discrepancy emerged in the PPV ($35.6 \pm 3.5\%$) compared to that of held-out data ($73.3 \pm 0.5\%$). This discrepancy was not observed in \texttt{Clinic} settings where participants were from the U.S. In addition, the model was performing significantly better among the female participants from Bangladesh, compared to the males. This is likely due to potential cultural differences in how people smile~\cite{fang2019unmasking}, compounded by the under-representation of individuals from Bangladesh in the training data. Specifically, although the web-based PARK tool used for collecting training data was intended to be used globally, the participants were predominantly U.S. residents ($754$ out of $827$, or $91.2\%$), potentially failing to capture cultural intricacies and variances in videos collected from Bangladesh.

On held-out data, our model did not demonstrate any detectable bias based on protected attributes such as ethnicity and sex. However, the model was less accurate for older participants. As most young participants in the dataset are healthy, the model could identify them as non-PD by indirectly inferring from attributes tied to being young. In addition, some subgroup-wise statistical analyses on the external test cohorts were less powerful (or infeasible) due to the subgroup population being small (or non-existent). This lack of representation from under-served demography or geographic regions is a limitation that should be addressed to strengthen the generalizability of the model. 
Our Spearman's correlation test found a moderate correlation coefficient ($\rho = -0.35$) between disease duration and error rate, suggesting that the model might be higher error-prone for groups with lower disease duration. However, this correlation was not statistically significant and the Chi-Squared test revealed no significant differences in error rates across the subgroups of different disease duration. It is worth noting that higher duration groups had significantly fewer participants, which might contribute to the observed performance deficiencies. We recommend that future studies enhance data diversity and representation within the training set, including across different disease durations, potentially via targeted recruitment, before deploying the model in diverse socio-cultural settings to ensure optimal performance.

While our study has yielded valuable insights, several areas present opportunities for further exploration. 
Firstly, our predictive model was evaluated on a single external clinical cohort and a cohort from a single low-income country. 
Expanding our validation by including data from multiple clinics and PD care facilities across diverse cultural and economic contexts could enhance the generalizability of our findings.
Secondly, hypomimia is typically assessed with natural facial expressions. Our participants were asked to smile in front of a camera (or mimic another expression), which may have resulted in unnatural and forced expressions. 
Future studies can probe more into investigating both natural and prompted expressions to explore the similarities and differences in their impact on assessment outcomes.
Moreover, it is important to note that not all individuals with PD exhibit every symptom. An individual with PD may have hand tremors but no facial muscle stiffness, and vice versa. Consequently, our algorithm may have limited utility for people who may not show initial symptoms through facial expressions. 
Finally, PD progression is not unimodal and can encompass different combinations of cognitive, motor, and behavioral changes~\cite{anderson2004behavioral,roheger2018progression,rektorova2019current,bugalho2021progression}. This multimodal nature of PD implies that a comprehensive screening approach should account for the diverse manifestations of the disease. By incorporating various modalities, such as speech analysis, motor coordination assessment, and cognitive evaluations, we can provide a more holistic understanding of an individual's condition.

\section*{Declaration of interests}
All the authors declare that there are no competing interests.

\section*{Data and code sharing}
To maintain compliance with the Health Insurance Portability and Accountability Act (HIPAA), we are unable to share the original videos, as they contain identifiable information such as participants' faces. However, we will publicly share the video processing codes, de-identified data containing the extracted features, and model training codes upon acceptance of the manuscript.

\section*{Contributors}
TA, MSI, WR, SL worked on data analysis, feature extraction, model training, model interpretation, and manuscript preparation. RBS, JLA and ERD led the data collection effort from the clinical cohort and helped to interpret the results and their potential clinical applications. SDT, KN, and IS lead the data collection efforts from Bangladesh, and performed preliminary data analysis. All authors contributed to writing this manuscript. EH was the principal investigator of the project; he facilitated the entire project and helped shape the narrative of the manuscript.

\section*{Acknowledgments}
We express our sincere gratitude to Cathe Schwartz, Karen Jaffe, and the dedicated members of the InMotion PD care facility for their invaluable assistance in collecting a diverse dataset from the residents of Ohio, United States. We would also like to extend our appreciation to Meghan Pawlik, Phillip Yang, and the team at the University of Rochester's Center for Health and Technology (CHeT). Their collaboration and support played a pivotal role in facilitating the data collection process from the clinical cohort. Finally, we would like to acknowledge the significant contributions of Dr. Amir Zadeh, Dr. M Rafayet Ali, Dr. Raiyan Abdul Baten, and Abdelrahman Abdelkader from the University of Rochester. Their diligent efforts in data collection and preliminary analysis laid the foundations for this work.

\section*{Use of Large Language Models} Sophisticated large language models (LLMs such as ChatGPT and Gemini) were used as editorial aid during the preparation of this manuscript. The primary use case of LLMs was to refine the manuscript's language and grammar through interactive feedback. All recommendations proposed by the LLMs were evaluated by an author prior to their final incorporation into the text. The usage of the LLMs was strictly limited to proposing amendments to the existing content. LLMs were not utilized to generate any new material.

\newpage
\section*{Supplementary Materials}
\captionsetup[figure]{name=Supplementary Figure}
\captionsetup[table]{name=Supplementary Table}

\section*{Supplementary Note 1 -- Participant Recruitment}
In this research, we gathered video data from four distinct settings, comprising i) home-recorded videos from global participants (\texttt{Home-Global}) -- primarily from North America; ii) videos recorded at a clinic at the University of Rochester Medical Center (\texttt{Clinic}); iii) videos recorded at a Parkinson's disease care facility in Ohio (\texttt{PD Care Facility}); and iv) home-recorded videos from Bangladesh, a country in Southeast Asia (\texttt{Home-BD}).
For data collection from the \texttt{Home-Global} cohort, we advertised the PARK tool using social media, emailed participants enrolled in a PD clinical study registry, and verbally reached out to subjects interested to contribute in PD research. Participants in the \texttt{Clinic} cohort are recruited via a clinical study conducted by the University of Rochester Medical Center, where recording the videos was optional. Some of the participants were supervised or assisted by clinical study team members during the recording process. The \texttt{PD Care Facility} consists of clients of InMotion (a PD care facility located in Ohio, United States) and their caregivers. Finally, for the \texttt{Home-BD} cohort, the subjects were recruited from local clinics through in-person outreach and the PARK tool was translated into their native language.  

\section*{Supplementary Note 2 -- Feature Extraction}
In our study, we extracted two categories of features from the facial expression videos collected from the participants -- facial action units using  OpenFace~\cite{baltrusaitis2018openface} and facial landmark features using MediaPipe~\cite{lugaresi2019mediapipe}. \\

\textbf{Facial Action Unit Features.} We leveraged the Facial Action Coding System (FACS), developed by Ekman et al.~\cite{ekman1978facial} to taxonomize human facial expressions. FACS describes facial expressions in terms of individual components of facial movement, or Facial Action Units (AUs). The AUs are associated with the muscle movements of the face and activation of a particular AU indicates the movement of a fixed set of facial muscles. In the literature, OpenFace has been proven to be very accurate in detecting the presence and intensity of these AUs~\cite{amos2016openface, baltrusaitis2018openface}. Utilizing Openface, we extracted the AU values for each frame of the three distinct facial expression videos from each participant. OpenFace software gives a binary activation ($0$ or $1$) and a raw magnitude (ranging $0$ to $5$) of each AU for each frame of a video that contains a human face. We evaluated three distinct statistical measures -- mean, variance, and entropy -- of the raw action unit when the corresponding action unit is active (i.e., the activation value is $1$). The variance signals the extent of facial muscle movement during a facial expression, while the mean indicates the average intensity of muscle engagement for each facial expression throughout the video frames. Conversely, entropy provides a measurement of unpredictability or randomness in the activation of facial muscles during expressions. Higher entropy often corresponds to more complex or varied facial movements~\cite{zurek2018complexity}.\\

\textbf{Facial Landmark Features.} In addition to the action unit features, following the prior literature~\cite{gomez2021improving}, we extracted facial attributes that simulate clinical assessments typically carried out in-person. 

\begin{itemize}
    \item \textit{Opening of the left and right eye}: Patients with PD may experience a decreased frequency of eye blinking. This can lead to more prolonged openings of the eyes, subtly altering the natural dynamics of facial expressions. By including the measurement of the degree of eye openings in our feature list, we aim to capture this nuanced change.
    
    \item \textit{Rising of the left and right eyebrows}: As a part of mimicking the surprise facial expression, the participants in our study were instructed to raise their eyebrows. However, in the case of Parkinson's patients, who may exhibit reduced facial expressiveness due to hypomimia, this eyebrow movement can be less pronounced. By taking into account the extent of eyebrow-raising, our analysis considers these subtle variations.
    
    \item \textit{Opening of the mouth}: As we have discussed earlier, one of the visible manifestations of PD can be an inability to fully close the mouth when the face is at rest, especially noticeable in the moderate to severe stages of the disease. We thus selected the extent of mouth opening as one of our target features as it can provide valuable insights into the degree of facial muscle control loss, a critical indicator of PD progression.
    
    \item \textit{Width of the mouth}: The width of the mouth during a smile can serve as an important indicator of facial muscle coordination and control. Given that Parkinson's patients often exhibit less expressive and smaller smiles -- known as the ``Parkinson's mask''~\cite{tickle2004practitioners} -- the width of the mouth during smiling can play a significant role in identifying the presence of the disease.

    \item \textit{Opening of the jaw}: The degree of jaw opening, often correlated with the symptom of masked faces, is another distinctive facial feature in Parkinson's disease. Patients may have difficulties fully controlling their jaw movement due to muscular rigidity, one of the primary motor symptoms of Parkinson's. Therefore, we chose to include the measurement of jaw opening in our feature set.
\end{itemize}

To compute these attributes, we used the face mesh solution of MediaPipe. MediaPipe, a product developed by Google, provides $478$ 3D facial landmarks from a video frame containing a face, using a lightweight machine learning model. It first detects a face in the image or video stream and then applies the face mesh model to estimate the facial landmarks. 
We already mentioned that we extracted varied set of Action Units varied across different facial expression videos. However, to prepare the facial landmark features, we extracted the same set of seven facial attribute features from each expression using Mediapipe.
This systematic approach ensures that each facial expression contributes an equivalent amount of information to the final feature space. Following the extraction, we performed a statistical aggregate on these features extracted at the frame level, computing mean, variance, and entropy. This process results in another set of $63$ digital features ($3$ facial expressions $\times$ $7$ facial attributes per expression $\times$ $3$ statistical aggregates per attribute), further enriching the feature space. A comprehensive list of the extracted features for each facial expression task is provided in Table \ref{tab:facial-expressions-and-action-units}. 

\begin{table}[t]
\centering
\caption{Action Units extracted by Openaface and Facial Attributes extracted by Mediapipe.}
\resizebox{0.5\columnwidth}{!}{
\begin{tabular}{lll}
\toprule
\textbf{Category} & \multicolumn{2}{c}{\textbf{Facial Featues}} \\
\midrule
\textbf{Facial Expression}& \textbf{Action Units} & \textbf{Description} \\
\midrule
\multirow{7}{*}{\textbf{Smile}} & 1 & Inner Brow Raiser\\
& 6 & Cheek Raiser\\
& 12 & Lip Corner Puller\\
& 14 & Dimpler\\
& 25 & Lips Part\\
& 26 & Jaw Drop\\
& 45 & Blink\\
\hline
\multirow{7}{*}{\textbf{Disgust}} & 4 & Brow Lowerer\\
& 7 & Lid Tightener\\
& 9 & Nose Wrinkler\\
& 10 & Upper Lip Raiser\\
& 25 & Lips Part\\
& 26 & Jaw Drop\\
& 45 & Blink\\
\hline
\multirow{7}{*}{\textbf{Surprise}} & 1 & Inner Brow Raiser\\
& 2 & Outer Brow Raiser\\
& 4 & Brow Lowerer\\
& 5 & Upper Lid Raiser\\
& 25 & Lips Part\\
& 26 & Jaw Drop\\
& 45 & Blink\\
\bottomrule
\multirow{7}{*}{\textbf{Mediapipe Attributes}} & \multicolumn{2}{c}{Right Eye Open} \\
& \multicolumn{2}{c}{Left Eye Open} \\
& \multicolumn{2}{c}{Right Eye Raised} \\
& \multicolumn{2}{c}{Left Eye Raised} \\
& \multicolumn{2}{c}{Mouth Open} \\
& \multicolumn{2}{c}{Mouth Width} \\
& \multicolumn{2}{c}{Jaw Open} \\

\bottomrule
\end{tabular}
}
\label{tab:facial-expressions-and-action-units}
\end{table}

\section*{Supplementary Note 3 -- Model Training, Inference, and Evaluation}
After combining the features derived from facial action units and landmarks extracted by MediaPipe, we had a $126$-dimensional feature space ($63$ from AU features and $63$ from facial landmark features) to represent all three facial expressions for each participant. When all three facial expressions were used to train predictive models, all of these 126 features were taken into consideration. However, we also investigated using only one facial expression or a different combination of two facial expressions. In those cases, we only used the corresponding features related to the facial expressions we are using. For example, only $42$ relevant features (i.e., one-third of all features) were considered when we trained the predictive model with merely the smile expression. The model training phase includes feature selection, feature scaling, creating synthetic samples for the minority class to account for class imbalance, and a training loop where the model gradually learned from data. During inference, we applied the same feature selection and scaling techniques and then used the learned model to make a prediction.\\

\textbf{Feature Selection.} Feature selection is a widely adopted approach aimed at reducing the dimensionality of data~\cite{li2017feature,zebari2020comprehensive}. By selecting a subset of relevant features, this technique helps prevent overfitting and enables the training of simpler and more generalizable models. We explored three different feature selection techniques: (i) logistic regression coefficients, (ii) Boosted Recursive Feature Elimination (BoostRFE), and (iii) Boosted Recursive Feature Addition (BoostRFA). Each of these methods ranks the features based on their importance in constructing a predictive model that can effectively differentiate participants with PD from those without the condition. After the features are ranked, we select top-$n$ features as input to the predictive model. Note that $n$ is considered a hyperparameter and tuned with other parameters in order to derive the best-performing model. 

For the logistic regression-based feature ranking, we ran a logistic regression on the full dataset (after scaling and normalizing the feature values), and used the absolute coefficient of each feature as a proxy of its importance~\cite{cheng2006logistic,zakharov2011ensemble}. Both BoostRFE and BoostRFA approach build an estimator (i.e., gradient boosting) with all available features and rank them based on feature importance obtained by SHapley Additive exPlanations (SHAP)~\cite{NIPS2017_7062}. BoostRFE iteratively removes the least important feature one by one, as long as the model's performance improves, until only the top-n features remained~\cite{saberi2022lightgbm,jeon2020hybrid}. On the other hand, BoostRFA starts with the most important feature and incrementally adds the next most important feature if it led to performance improvement~\cite{hamed2018network}. The process is continued until $n$ features are added to the model.

In our study, we found that feature selection, in general, led to performance improvement, and the logistic regression-based feature ranking outperformed the other two approaches (see Table \ref{tab:results}). Therefore, the logistic-regression-based feature selection was used as the default for other experiments.\\

\textbf{Feature Scaling.} Feature scaling is a crucial pre-processing step in machine learning~\cite{sharma2022study,ahsan2021effect}. It ensures that all features are on a similar scale or range, and often leads to performance improvement. We have tried two different feature scaling methods mentioned below:

\begin{itemize}
    \item \textbf{MinMax Scaling} scales the values of each feature between 0 and 1 using the following formula:

    $$X_{\text{scaled}} = \frac{(X - X_{\min})}{(X_{\max}-X_{\min})}$$

    where $X$ is the original feature value, and $X_{\min}, X_{\max}$ are the minimum and the maximum value of the feature, respectively, in the entire dataset.

    \item \textbf{Standard Scaling} ensures the values of the input features are scaled to have a mean of 0 and a standard deviation of 1. The formula for Standard Scaling is:
    $$X_{\text{scaled}} = \frac{X - \mu}{\sigma}$$
    
    where $X$ is the original feature value, and $\mu, \sigma$ are the mean and standard deviation of the feature, respectively, in the entire dataset.
\end{itemize}

Based on ablation studies (Table \ref{tab:results}), feature scaling helped boost the performance of the predictive model. Among the two methods, we select the MinMax Scaler as the default scaling method as it yielded the best performance.\\

\textbf{Minority Oversampling.} The dataset we studied was imbalanced, as it contained $803$ participants without PD compared to $256$ with the condition. Such class imbalance poses a challenge for training machine learning models as it can lead to a biased model that might overfit the majority class and under-represent the minority class, resulting in poor generalization. To address this issue, we employed the Synthetic Minority Oversampling Technique (SMOTE)~\cite{chawla2002smote}. SMOTE is a widely-used approach to balance class distribution through the generation of synthetic instances of the minority class. As shown in Table \ref{tab:results}, incorporating SMOTE in model training helped improve the performance of the predictive model. However, please note that the entire dataset was split into train and test sets first, and then SMOTE was applied only on the train set. This ensures that the test set only contained real data, and none of the synthetic data in the training set was derived from any test data.\\

\textbf{Evaluation.} We used $k$-fold cross-validation with stratified sampling to assess the performance of the predictive model on held-out data. This approach involves dividing the dataset into $k$ different folds, with each fold containing $\frac{1}{k}$ portion of the dataset. The key distinction of stratified sampling is that it ensures each fold maintains a proportional representation of samples from different classes, mirroring the distribution in the entire dataset. To illustrate, consider an example where the dataset consists of $100$ samples, with $10$ samples belonging to class I and $90$ samples belonging to class II ($1:9$ ratio). In $10$-fold stratified cross-validation, each fold would include $1$ samples from class I and $9$ samples from class II, preserving the ratio of samples per class observed in the full dataset. This ensures that the model is trained and evaluated on diverse samples from all classes. The evaluation process encompasses $k$ iterations, with each iteration reserving one of the folds as the test set, while the model is trained on the remaining data. This procedure allows for a comprehensive assessment of the model's performance across the entire dataset while addressing potential biases introduced by class imbalances. 

In addition to the held-out data, we used two different external test sets collected in (i) \texttt{Clinic} and (ii) \texttt{Home-BD} settings. For evaluating our model on these test sets, the model was solely trained on all data collected in the other two settings (i.e., \texttt{Home-Global} and \texttt{PD Care Facility}). As a result, the model has never seen any data from either \texttt{Clinic} or \texttt{Home-BD} setting. For each of the external test sets, we selected and scaled the features according to the training process, and ran inference using the trained model.

We used a comprehensive set of evaluation metrics widely adopted by machine learning for the healthcare community: (i) Area Under the Receiver Operating Characteristic curve (AUROC), (ii) binary Accuracy, (iii) Sensitivity, (iv) Specificity, (v) Positive Predictive Value (PPV), and (vi) Negative Predictive Value (NPV). Instead of relying on a single metric, we selected the best models (reported in this paper) based on a holistic evaluation of all of the above metrics. 
For automatically selecting the top-performing models as input to the final ensemble model, we used the models with the highest AUROC scores.\\

\definecolor{green1}{HTML}{036D15}
\definecolor{green2}{HTML}{05A921}
\definecolor{green3}{HTML}{009900}
\definecolor{green4}{HTML}{05EC2B}

\renewcommand{\arraystretch}{1.22}

\begin{table*}[!htbp]
\centering
\caption{\textbf{Performance reporting for the ablation studies.}}
\resizebox{1.0\linewidth}{!}{%
\tiny
\begin{tabular}{rrrrrrr}
\arrayrulecolor{green2}\hline
\multicolumn{1}{l}{\textcolor{green1}{\textbf{Experimental Setup}}} & \multicolumn{1}{l}{\textcolor{green1}{\textbf{AUROC}}} & \multicolumn{1}{l}{\textcolor{green1}{\textbf{Accuracy}}} & \multicolumn{1}{l}{\textcolor{green1}{\textbf{Sensitivity}}} & \multicolumn{1}{l}{\textcolor{green1}{\textbf{Specificity}}} & \multicolumn{1}{l}{\textcolor{green1}{\textbf{PPV}}} & \multicolumn{1}{l}{\textcolor{green1}{\textbf{NPV}}} \\
\arrayrulecolor{green2}\hline
\arrayrulecolor{green2}\hline
\\[-2.5ex]
\arrayrulecolor{green4}\hline
\multicolumn{7}{l}{\textcolor{green3}{\textbf{1. Machine Learning Baselines}}} \\
\arrayrulecolor{green4}\hline

\multicolumn{1}{l}{XGBoost} & $89.2$ & $84.5$ & $75.4$ & $87.3$ & $64.8$ & $92.0$ \\
\multicolumn{1}{l}{LightGBM} & $89.0$ & $85.5$ & $71.8$ & $89.7$ & $68.3$ & $91.2$ \\
\multicolumn{1}{l}{Random Forest} & $88.5$ & $83.1$ & $75.9$ & $85.3$ & $61.4$ & $92.0$ \\
\multicolumn{1}{l}{AdaBoost} & $89.3$ & $85.1$ & $76.9$ & $87.7$ & $65.8$ & $92.5$ \\
\multicolumn{1}{l}{SVM} & $87.9$ & $86.2$ & $78.5$ & $88.6$ & $68.0$ & $93.0$ \\
\multicolumn{1}{l}{Hist Gradient Boosting} & $90.8$ & $86.8$ & $74.4$ & $90.7$ & $71.1$ & $92.0$ \\
\arrayrulecolor{gray}\hline
\multicolumn{1}{l}{Ensemble of best models with SVM} & $88.5$ & $88.0$ & $76.9$ & $91.5$ & $73.5$ & $92.8$ \\
\multicolumn{1}{l}{\textbf{Ensemble of best models with LR}} & $\mathbf{90.1}$ & $\mathbf{88.0}$ & $\mathbf{76.4}$ & $\mathbf{91.6}$ & $\mathbf{73.8}$ & $\mathbf{92.6}$ \\
\arrayrulecolor{green4}\hline
\multicolumn{7}{l}{\textcolor{green3}{\textbf{2. Feature Scaling Methods}}} \\
\arrayrulecolor{green4}\hline
\multicolumn{1}{l}{No Scaler} & $88.8$ & $86.8$ & $75.9$ & $90.2$ & $70.5$ & $92.4$ \\
\multicolumn{1}{l}{Standard Scaler} & $89.2$ & $87.2$ & $78.0$ & $90.0$ & $70.7$ & $93.0$ \\
\multicolumn{1}{l}{\textbf{MinMax Scaler}} & $\mathbf{90.1}$ & $\mathbf{88.0}$ & $\mathbf{76.4}$ & $\mathbf{91.6}$ & $\mathbf{73.8}$ & $\mathbf{92.6}$ \\
\arrayrulecolor{green4}\hline
\multicolumn{7}{l}{\textcolor{green3}{\textbf{3. Feature Selection Methods}}} \\
\arrayrulecolor{green4}\hline
\multicolumn{1}{l}{No Feature Selection} & $87.1$ & $87.9$ & $70.8$ & $93.2$ & $76.2$ & $91.2$ \\
\multicolumn{1}{l}{BoostRFA} & $87.0$ & $87.3$ & $76.9$ & $90.5$ & $71.4$ & $92.7$ \\
\multicolumn{1}{l}{BoostRFE} & $87.4$ & $87.6$ & $78.5$ & $90.4$ & $71.5$ & $93.2$ \\
\multicolumn{1}{l}{\textbf{Logistic Regression}} & $\mathbf{90.1}$ & $\mathbf{88.0}$ & $\mathbf{76.4}$ & $\mathbf{91.6}$ & $\mathbf{73.8}$ & $\mathbf{92.6}$ \\
\arrayrulecolor{green4}\hline
\multicolumn{7}{l}{\textcolor{green3}{\textbf{4. Combination of Facial Expressions}}} \\ 
\arrayrulecolor{green4}\hline
\multicolumn{1}{l}{\textbf{Smile}} & $\mathbf{90.1}$ & $\mathbf{88.0}$ & $\mathbf{76.4}$ & $\mathbf{91.6}$ & $\mathbf{73.8}$ & $\mathbf{92.6}$ \\
\multicolumn{1}{l}{Disgust} & $76.2$ & $77.5$ & $69.7$ & $79.9$ & $51.7$ & $89.5$ \\
\multicolumn{1}{l}{Surprise} & $72.5$ & $77.0$ & $61.0$ & $82.0$ & $51.1$ & $87.2$ \\
\multicolumn{1}{l}{Smile + Disgust} & $87.6$ & $86.3$ & $78.0$ & $88.9$ & $68.5$ & $92.9$ \\
\multicolumn{1}{l}{Smile + Surprise} & $87.4$ & $88.0$ & $78.5$ & $91.0$ & $72.9$ & $93.2$ \\
\multicolumn{1}{l}{Disgust + Surprise} & $72.5$ & $76.1$ & $66.2$ & $79.1$ & $49.4$ & $88.3$ \\
\multicolumn{1}{l}{\textbf{Smile + Disgust + Surprise}} & $\mathbf{89.3}$ & $\mathbf{89.7}$ & $\mathbf{76.9}$ & $\mathbf{93.7}$ & $\mathbf{79.0}$ & $\mathbf{92.9}$ \\
\arrayrulecolor{green4}\hline
\multicolumn{7}{l}{\textcolor{green3}{\textbf{5. Combination of AU and Landmark Features}}} \\ 
\arrayrulecolor{green4}\hline
\multicolumn{1}{l}{AU Features} & $88.1$ & $85.5$ & $69.2$ & $90.5$ & $69.2$ & $90.5$ \\
\multicolumn{1}{l}{Landmark Features} & $71.1$ & $77.6$ & $48.7$ & $86.6$ & $52.8$ & $84.5$ \\
\multicolumn{1}{l}{\textbf{AU + Landmark Features}} & $\mathbf{90.1}$ & $\mathbf{88.0}$ & $\mathbf{76.4}$ & $\mathbf{91.6}$ & $\mathbf{73.8}$ & $\mathbf{92.6}$ \\
\arrayrulecolor{green4}\hline
\multicolumn{7}{l}{\textcolor{green3}{\textbf{6. Impact of Minority Oversampling}}} \\ 
\arrayrulecolor{green4}\hline
\multicolumn{1}{l}{Not Used} & $87.1$ & $86.7$ & $63.1$ & $94.0$ & $76.4$ & $89.2$ \\
\multicolumn{1}{l}{\textbf{Used}} & $\mathbf{90.1}$ & $\mathbf{88.0}$ & $\mathbf{76.4}$ & $\mathbf{91.6}$ & $\mathbf{73.8}$ & $\mathbf{92.6}$ \\
\arrayrulecolor{green4}\hline
\multicolumn{7}{l}{\textcolor{green3}{\textbf{7. Evaluation on External Data}}} \\
\arrayrulecolor{green4}\hline
\multicolumn{1}{l}{Clinic Cohort} & $82.0$ & $80.0$ & $85.1$ & $71.4$ & $83.3$ & $74.1$ \\
\multicolumn{1}{l}{{Home-BD Cohort}} & {$81.5$} & {$85.2$} & {$71.4$} & {$86.7$} & {$35.7$} & {$96.7$} \\
\arrayrulecolor{green4}\hline
\end{tabular}

}
\label{tab:results}
\end{table*}

\textbf{Hyper Parameter Tuning.} Parameter tuning~\cite{victoria2021automatic,feurer2019hyperparameter} plays a crucial role in enhancing the performance of machine learning models. It involves adjusting the hyperparameters of these models to optimize their predictive performance. For this study, we employed a range of classifiers, including Support Vector Machine (SVM), AdaBoost, HistBoost, XGBoost, and Random Forest. The hyperparameters of these classifiers were tuned using Weights \& Biases (\textit{WandB})~\cite{wandb}, a machine learning tool designed for this specific purpose. \textit{WandB} helped us identify the optimal combination of parameters that yielded the best model performance. The core principle underlying hyperparameter tuning is to traverse the search space of various parameter combinations with the goal of maximizing or minimizing a specific performance metric, for which, we used a Bayesian approach enabled by WandB.

For example, in our study, the parameters tuned for the SVM model included the number of top significant features (\textit{n}), the penalty parameter of the error term (\textit{C}), and the kernel coefficient for `rbf', `polynomial' and `sigmoid' kernels (\textit{gamma}). For AdaBoost, apart from \textit{n}, the number of weak learners (\textit{n\_estimators}) and learning rate were tuned. Similarly, for the XGBoost and HistBoost models, we tuned parameters such as maximum depth of a tree (\textit{max\_depth}), minimum sum of instance weight needed in a child (\textit{min\_child\_weight}), and boosting learning rate (\textit{eta}), among others. In the Random Forest classifier, the number of trees in the forest (\textit{n\_estimators}) and the maximum depth of the tree (\textit{max\_depth}) were the key parameters tuned. The objective of Bayesian hyper-parameter tuning approach was to maximize AUROC. The scripts for hyper-parameter tuning will be released alongside the code upon acceptance of the manuscript.\\

\textbf{Ensemble Modeling.} We designed the ensemble modeling~\cite{dong2020survey,dietterich2002ensemble,sagi2018ensemble} approach to consolidate the strengths of {Histogram-Based Gradient Boosting (HistBoost)} models to achieve better predictive performance than any individual model. By leveraging the diversity introduced via different feature sets and model parameters, the ensemble model aimed to reduce variance and potentially offer a more robust decision boundary for classification. We first trained individual HistBoost models as mentioned above. After that, we selected the best-performing $m$ (a hyper-parameter) HistBoost models based on the AUROC scores. The predictions of these chosen individual HistBoost models served as the ensemble's meta-features. These meta-features were fed into a second layer model, which acted as the final classifier. In our exploration for the optimal second-layer model, we considered both an HistBoost and a logistic regression classifier.

It's essential to underscore that introducing this second layer wasn't a mere procedural step. Both the HistBoost and the logistic regression models underwent rigorous hyper-parameter tuning, ensuring they were precisely calibrated for the ensemble. Eventually, we finalized on the logistic regression classifier as the second layer model as it came out as a superior choice among the two with enhanced performance metrics. This additional layer of modeling aimed to capture any complementary decision patterns that might be present among the individual HistBoost models. Ensemble methods, in general, exploit the idea that a group of `weak learners' can come together to form a `strong learner'. In our case, while each HistBoost was already a strong learner in its own right, by combining their predictions through a logistic regression meta-classifier, we aimed to further enhance the predictive power and generalization of the final model. By doing so, the ensemble model is expected to mitigate individual model biases, reduce overfitting, and potentially lead to better performance on unseen data.\\

\textbf{Ablation Study.} The best models were chosen based on the cross-validation performance obtained from different model choices, including machine learning baselines, feature scaling and selection methods, combination of facial expressions, combination of action unit and landmark features, and whether to oversample the minority class. The results of this ablation study is summarized in Table \ref{tab:results}.\\

\textbf{SHAP Analysis.}
SHAP (SHapley Additive exPlanations) analysis is a unified approach to interpreting the output of machine learning models by assigning each feature an importance value based on Shapley values from cooperative game theory~\cite{NIPS2017_7062}. This method considers all possible combinations of features, ensuring a fair distribution of contributions among them. SHAP values provide a consistent and interpretable measure of feature impact, which is crucial for understanding model behavior, especially in complex, non-linear models. By quantifying the contribution of each feature to the prediction, SHAP analysis enhances transparency and trust in the model's decisions, enabling the identification of key drivers of the model's outputs and ensuring that the model's predictions align with domain knowledge and clinical relevance.

\section*{Supplementary Note 4 -- Additional Plots for Statistical Test on External Validation}
We performed a series of statistical tests to ensure that our best-performing model delivers equitable performance across different demographic subgroups. The model was validated on two external datasets: the \texttt{Clinic} cohort (collected from a clinic in New York) and the \texttt{Home-BD} cohort (collected from Bangladesh). We assessed the misclassification rates across male and female subgroups, as well as across various age groups. Since the majority of participants in the \texttt{Clinic} cohort were white and all \texttt{Home-BD} participants were Asian, ethnicity-based analysis was not conducted. Supplementary Figures~\ref{fig:external_sex} and \ref{fig:external_age} present the results for sex and age, respectively.

\begin{figure}[t]
  \centering
  \begin{subfigure}[t]{.48\textwidth}
    \centering
    \includegraphics[width=1.0\linewidth]{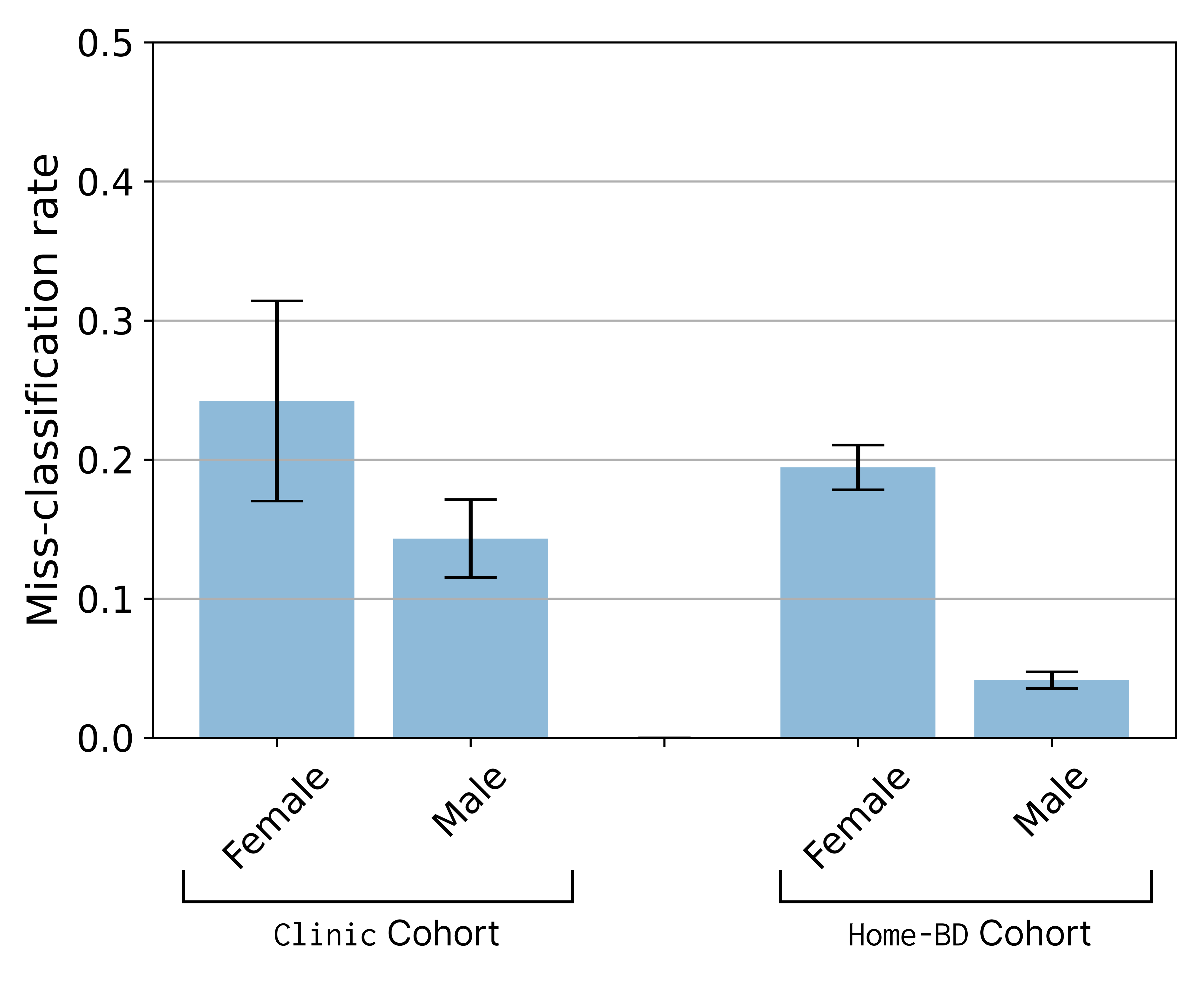}
    \caption{}
    \label{fig:external_sex}
  \end{subfigure}
  \hfill
  \begin{subfigure}[t]{.48\textwidth}
    \centering
    \includegraphics[width=1.0\linewidth]{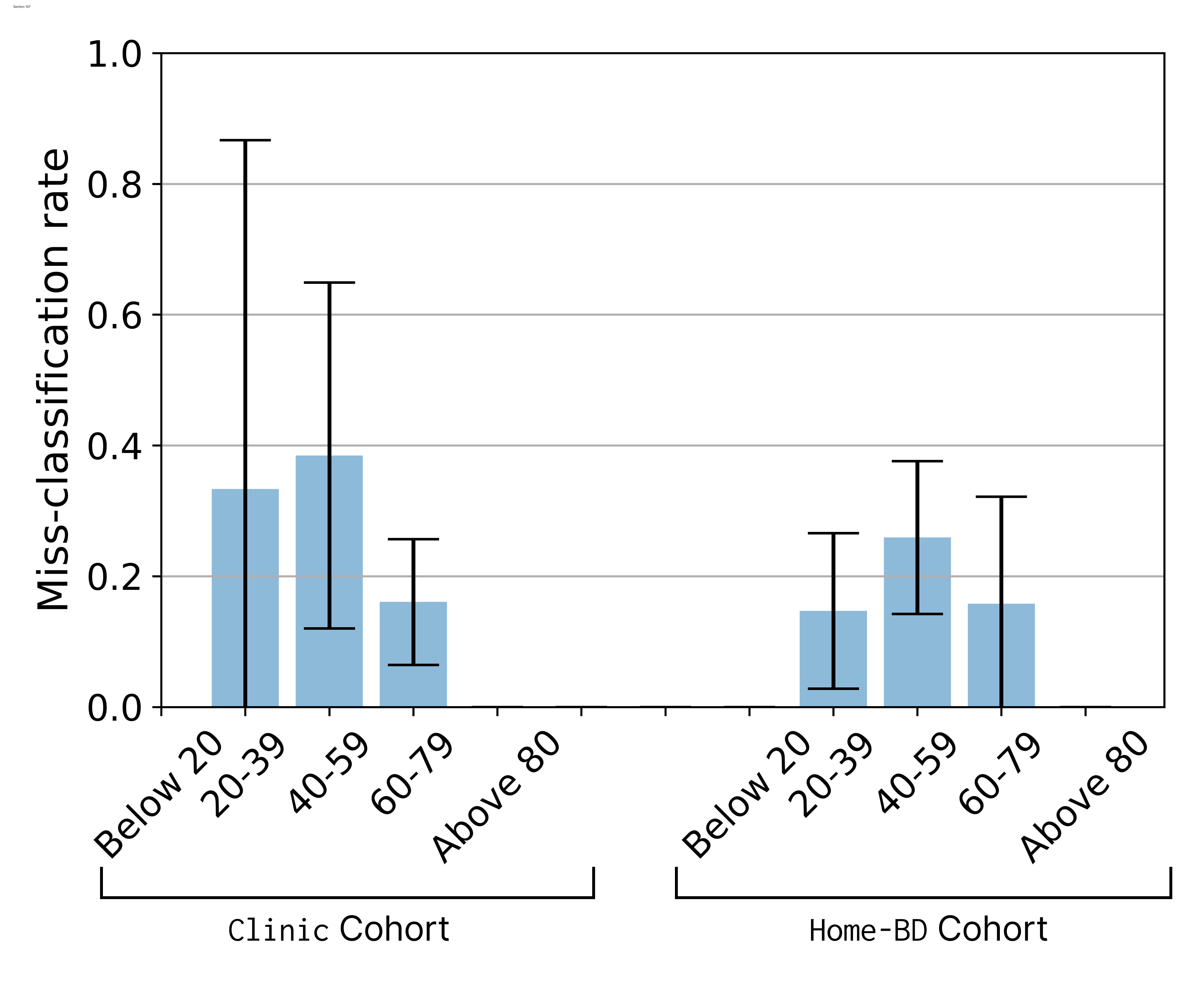}
    \caption{}
    \label{fig:external_age}
  \end{subfigure}

  \caption{\textbf{Comparison of model performance across sex and age subgroups.} The first plot shows the misclassification rates for male and female subgroups in both the \texttt{Clinic} and \texttt{Home-BD} cohorts. The second plot compares performance across different age groups. In all cases, The error bars represent $95\%$ confidence interval.}  
  \label{fig:comprehensive_analysis}
\end{figure}

\newpage
\section*{Supplementary Note 5 -- Demographic Characteristics of Different Cohorts}
In this section, we provide detailed demographic information for data collected from four different cohorts mentioned in the manuscript. Table \ref{tab:demographics-home-global}, \ref{tab:demographics-pd-care-facility}, \ref{tab:demographics-clinic}, and 
\ref{tab:demographics-home-bd} describe the details for \texttt{Home-Global}, \texttt{PD Care Facility}, \texttt{Home-BD}, and \texttt{Clinic} cohorts, respectively.

\begin{table}[!htbp]
\centering
\renewcommand{\arraystretch}{1.2} 

\begin{minipage}[t]{0.48\textwidth}
    \centering
    \caption{\textbf{Demographic information of the participants from \texttt{Home-Global} cohort.}}
    \resizebox{\columnwidth}{!}{%
    \begin{tabular}{lllll}
    \toprule
    \multicolumn{2}{l}{\textbf{Characteristics}} & \textbf{With PD} & \textbf{Without PD} & \textbf{Total} \\ \hline
    \multicolumn{2}{l}{\begin{tabular}[c]{@{}l@{}}Number of \\ Participants, \\ n (\%) \end{tabular} } & $77 (11.11\%)$ & $616 (88.89\%)$ & \textbf{693 (100\%)} \\ \hline
    \multicolumn{5}{l}{Sex, n (\%)} \\
    & Male & $48 (62.34\%)$ & $245 (39.77\%)$ & $\mathbf{293 (42.28\%)}$ \\ 
    & Female & $29 (37.66\%)$ & $371 (60.23\%)$ & $\mathbf{400 (57.72\%)}$ \\ 
    \midrule
    \multicolumn{5}{l}{Age in years (range: 18 -- 89, mean: 60.28), n (\%)} \\
    & \textless{}20 & $0 (0.0\%)$ & $6 (0.97\%)$ & $\mathbf{6 (0.87\%)}$ \\
    & 20-39 & $3 (3.9\%)$ & $51 (8.28\%)$ & $\mathbf{54 (7.79\%)}$ \\
    & 40-59 & $15 (19.48\%)$ & $159 (25.81\%)$ & $\mathbf{174 (25.11\%)}$ \\
    & 60-79 & $58 (75.32\%)$ & $397 (64.45\%)$ & $\mathbf{455 (65.66\%)}$ \\
    & \textgreater{}=80 & $1 (1.3\%)$ & $3 (0.49\%)$ & $\mathbf{4 (0.58\%)}$ \\
    \midrule
    \multicolumn{5}{l}{Race, n (\%)} \\
     & White  & $70 (90.91\%)$ & $504 (81.82\%)$ & $\mathbf{574 (82.83\%)}$ \\
     & \textcolor{blue}{Asian}  & $1 (1.3\%)$ & $52 (8.44\%)$ & $\mathbf{53 (7.65\%)}$ \\
     & \begin{tabular}[c]{@{}l@{}}Black or \\ African \\ American\end{tabular} & $3 (3.9\%)$ & $44 (7.14\%)$ & $\mathbf{47 (6.78\%)}$ \\
     & \begin{tabular}[c]{@{}l@{}}\textcolor{blue}{American} \\ \textcolor{blue}{Indian or} \\ \textcolor{blue}{Alaska} \\ \textcolor{blue}{Native}\end{tabular}  & $0 (0.0\%)$ & $4 (0.65\%)$ & $\mathbf{4 (0.58\%)}$ \\
    & Others  & $3 (3.9\%)$ & $12 (1.95\%)$ & $\mathbf{15 (2.16\%)}$ \\
    & \textcolor{blue}{Not Mentioned}  & $0 (0.0\%)$ & $0 (0.0\%)$ & $\mathbf{0 (0.0\%)}$ \\
    \bottomrule
    \end{tabular}%
    }
    \label{tab:demographics-home-global}
\end{minipage}%
\hfill
\begin{minipage}[t]{0.48\textwidth}
    \centering
    \caption{\textbf{Demographic information of the participants from \texttt{PD Care Facility} cohort.} The policy of the care facility did not allow us to collect information related to participants' ethnicity.}
    \resizebox{\columnwidth}{!}{%
    \begin{tabular}{lllll}
    \toprule
    \multicolumn{2}{l}{\textbf{Characteristics}} & \textbf{With PD} & \textbf{Without PD} & \textbf{Total} \\ \hline
    \multicolumn{2}{l}{\begin{tabular}[c]{@{}l@{}}Number of \\ Participants, \\ n (\%) \end{tabular} } & $118 (83.1\%)$ & $24 (16.9\%)$ & \textbf{142 (100\%)} \\ \hline
    \multicolumn{5}{l}{Sex, n (\%)} \\
    & Male & $60 (50.85\%)$ & $12 (50.0\%)$ & $\mathbf{72 (50.7\%)}$ \\ 
    & Female & $58 (49.15\%)$ & $12 (50.0\%)$ & $\mathbf{70 (49.3\%)}$ \\ 
    \midrule
    \multicolumn{5}{l}{Age in years (range: 18 -- 93, mean: 67.27), n (\%)} \\
    & \textless{}20 & $0 (0.0\%)$ & $0 (0.0\%)$ & $\mathbf{0 (0.0\%)}$ \\
    & 20-39 & $1 (0.85\%)$ & $8 (33.33\%)$ & $\mathbf{9 (6.34\%)}$ \\
    & 40-59 & $11 (9.32\%)$ & $3 (12.5\%)$ & $\mathbf{14 (9.86\%)}$ \\
    & 60-79 & $95 (80.51\%)$ & $11 (45.83\%)$ & $\mathbf{106 (74.65\%)}$ \\
    & \textgreater{}=80 & $11 (9.32\%)$ & $2 (8.33\%)$ & $\mathbf{13 (9.15\%)}$ \\
    \bottomrule
    \end{tabular}%
    }
    \label{tab:demographics-pd-care-facility}
\end{minipage}

\end{table}

\begin{table}[!htbp]
\centering
\begin{minipage}[t]{.48\textwidth}
\centering
\caption{\textbf{Demographic information of the participants from \texttt{Clinic} cohort.} The clinic is located in Rochester, New York, where the majority race is white.}
\resizebox{\textwidth}{!}{%
\begin{tabular}{lllll}
\toprule
\multicolumn{2}{l}{\textbf{Characteristics}} & \textbf{With PD} & \textbf{Without PD} & \textbf{Total} \\ \hline
\multicolumn{2}{l}{\begin{tabular}[c]{@{}l@{}}Number of \\ Participants, \\ n (\%) \end{tabular} } & $47 (62.67\%)$ & $28 (37.33\%)$ & \textbf{75 (100\%)} \\ \hline
\multicolumn{5}{l}{Sex, n (\%)} \\
& Male & $31 (65.96\%)$ & $12 (42.86\%)$ & $\mathbf{43 (57.33\%)}$ \\ 
& Female & $16 (34.04\%)$ & $16 (57.14\%)$ & $\mathbf{32 (42.67\%)}$ \\ 
\midrule
\multicolumn{5}{l}{Age in years (range: 18 -- 86, mean: 65.43), n (\%)} \\
& \textless{}20 & $0 (0.0\%)$ & $0 (0.0\%)$ & $\mathbf{0 (0.0\%)}$ \\
& 20-39 & $0 (0.0\%)$ & $3 (10.71\%)$ & $\mathbf{3 (4.0\%)}$ \\
& 40-59 & $7 (14.89\%)$ & $6 (21.43\%)$ & $\mathbf{13 (17.33\%)}$ \\
& 60-79 & $38 (80.85\%)$ & $18 (64.29\%)$ & $\mathbf{56 (74.67\%)}$ \\
& \textgreater{}=80 & $2 (4.26\%)$ & $1 (3.57\%)$ & $\mathbf{3 (4.0\%)}$ \\
\midrule
\multicolumn{5}{l}{Race, n (\%)} \\
 & White  & $46 (97.87\%)$ & $27 (96.43\%)$ & $\mathbf{73 (97.33\%)}$ \\
 & \textcolor{blue}{Asian}  & $0 (0.0\%)$ & $0 (0.0\%)$ & $\mathbf{0 (0.0\%)}$ \\
 & \begin{tabular}[c]{@{}l@{}}Black or \\ African \\ American\end{tabular} & $0 (0.0\%)$ & $1 (3.57\%)$ & $\mathbf{1 (1.33\%)}$ \\
 & \begin{tabular}[c]{@{}l@{}}\textcolor{blue}{American} \\ \textcolor{blue}{Indian or} \\ \textcolor{blue}{Alaska} \\ \textcolor{blue}{Native}\end{tabular}  & $1 (2.13\%)$ & $0 (0.0\%)$ & $\mathbf{1 (1.33\%)}$ \\
& Others  & $0 (0.0\%)$ & $0 (0.0\%)$ & $\mathbf{0 (0.0\%)}$ \\
& \textcolor{blue}{Not Mentioned}  & $0 (0.0\%)$ & $0 (0.0\%)$ & $\mathbf{0 (0.0\%)}$ \\
\bottomrule
\end{tabular}%
}
\label{tab:demographics-clinic}
\end{minipage}%
\hfill
\begin{minipage}[t]{.48\textwidth}
\centering
\caption{\textbf{Demographic information of the participants from \texttt{Home-BD} cohort.} All the participants identified themselves as Asian (i.e., Bangladeshi).
}
\resizebox{\textwidth}{!}{%
\begin{tabular}{lllll}
\toprule
\multicolumn{2}{l}{\textbf{Characteristics}} & \textbf{With PD} & \textbf{Without PD} & \textbf{Total} \\ \hline
\multicolumn{2}{l}{\begin{tabular}[c]{@{}l@{}}Number of \\ Participants, \\ n (\%) \end{tabular} } & $14 (9.4\%)$ & $135 (90.6\%)$ & \textbf{149 (100\%)} \\ \hline
\multicolumn{5}{l}{Sex, n (\%)} \\
& Male & $11 (78.57\%)$ & $92 (68.15\%)$ & $\mathbf{103 (69.13\%)}$ \\ 
& Female & $3 (21.43\%)$ & $43 (31.85\%)$ & $\mathbf{46 (30.87\%)}$ \\ 
\midrule
\multicolumn{5}{l}{Age in years (range: 18 -- 80, mean: 41.18), n (\%)} \\
& \textless{}20 & $0 (0.0\%)$ & $37 (27.41\%)$ & $\mathbf{37 (24.83\%)}$ \\
& 20-39 & $0 (0.0\%)$ & $34 (25.19\%)$ & $\mathbf{34 (22.82\%)}$ \\
& 40-59 & $6 (42.86\%)$ & $48 (35.56\%)$ & $\mathbf{54 (36.24\%)}$ \\
& 60-79 & $7 (50.0\%)$ & $12 (8.89\%)$ & $\mathbf{19 (12.75\%)}$ \\
& \textgreater{}=80 & $1 (7.14\%)$ & $4 (2.96\%)$ & $\mathbf{5 (3.36\%)}$ \\
\bottomrule
\end{tabular}%
}
\label{tab:demographics-home-bd}
\end{minipage}
\end{table}

\bibliography{sample}

\end{document}